\documentclass[11pt]{article}

\usepackage{cite,latexsym,epsfig,euscript,amssymb,amsmath,verbatim,color,graphicx,multirow,hyperref,wrapfig,subfigure}
\usepackage[labelfont=bf,font={sl,small}]{caption}

\newcommand{\beq}{\begin{equation}}
\newcommand{\eeq}{\end{equation}}
\newcommand{\beqs}{\begin{eqnarray}}
\newcommand{\eeqs}{\end{eqnarray}}
\newcommand{\bit}{\begin{itemize}}
\newcommand{\eit}{\end{itemize}}
\newcommand{\bce}{\begin{center}}
\newcommand{\ece}{\end{center}}
\newcommand{\ben}{\begin{enumerate}}
\newcommand{\een}{\end{enumerate}}

\newcommand{\pa}{\mathcal{P}}

\newcommand{\nn}{\nonumber}
\newcommand{\ve}{\varepsilon}

\newcommand{\vegamma}{\varepsilon^{(\gamma)}}
\newcommand{\veg}{\varepsilon^{(g)}}
\newcommand{\eps}{\epsilon}
\newcommand{\gfive}{\gamma^5}
\newcommand{\half}{\frac{1}{2}}
\newcommand{\threehalf}{\frac{3}{2}}

\begin{document}

\title{
\vspace{2.5cm} 
\Large{\textbf{Jet-associated resonance spectroscopy}}\\[0.2cm]
}

\author{Christoph Englert$^{1}$, Gabriele Ferretti$^{2}$ and Michael Spannowsky$^{3}$ \\[2ex]
\small{\em $^1$SUPA, School of Physics and Astronomy, University of
  Glasgow,}\\
\small{\em Glasgow G12 8QQ, United Kingdom}\\[0.8ex]
\small{\em $^2$Department of Physics, Chalmers University of Technology,}\\
\small{\em Fysikg{\aa}rden, 41296 G\"oteborg, Sweden}\\[0.8ex] 
\small{\em $^3$Institute for Particle Physics Phenomenology, Department of Physics,} \\
\small{\em Durham University, Durham DH1 3LE, UK}\\[0.8ex] 
}

\date{}
\maketitle

\begin{abstract}
\noindent We present a model-independent study aimed at characterizing the nature of possible resonances in the jet-photon or jet-$Z$ final state at hadron colliders.
Such resonances are expected in many models of compositeness and would be a clear indication of new physics.
At leading order, in the narrow width approximation, the matrix elements are parameterized by just a few constants describing the coupling of the various helicities to the resonance. We present the full structure of such amplitudes up to spin two and use them to simulate relevant kinematic distributions that could serve to constrain the coupling structure. This also generalizes the signal generation strategy that is currently pursued by ATLAS and CMS to the most general case in the considered channels. While the determination of the P/CP properties of the interaction seems to be out of reach within this framework, there is a wealth of information to be gained about the spin of the resonance and the relative couplings of the helicities.
\end{abstract}
\vfill
\hfill{IPPP/17/49}

\thispagestyle{empty}

\newpage

\setcounter{page}{1} \pagestyle{plain} \renewcommand{\thefootnote}{\arabic{footnote}} \setcounter{footnote}{0}

%
%
\section{Introduction}

Many scenarios of dynamics Beyond the Standard Model (BSM), built with the aim to ameliorate the hierarchy problem, predict the existence of new resonances at the TeV scale. Examples of such states abound in many different contexts such as vector-like confinement~\cite{Kilic:2009mi,Schumann:2011ji}, compositeness~\cite{Hayot:1980gg,Ellis:1980hz,Belyaev:1999xe,Bai:2010mn,Cacciapaglia:2015eqa,Bai:2016czm,Bizot:2016zyu,Hayreter:2017wra}, partial compositeness~\cite{Barnard:2013zea,Ferretti:2013kya,Ferretti:2014qta,Vecchi:2015fma,Ferretti:2016upr,Belyaev:2016ftv} and excited quarks~\cite{Cabibbo:1983bk,DeRujula:1983ak,Kleiss:1987ab,Baur:1987ga,Baur:1989kv,Bhattacharya:2009xg}.

If such states are indeed observed in the future at the LHC (or at a future hadron collider), the first step in obtaining a detailed picture of the underlying theory is a dedicated spectroscopy program targeting the nature of their couplings, spin and CP properties.

These new resonances do not necessarily have to be QCD singlets and, thus, the general spectroscopy program that has been pursued in conjunction to the Higgs discovery~\cite{Choi:2002jk,Gao:2010qx,DeRujula:2010ys,Bolognesi:2012mm,Khachatryan:2014kca,Miller:2001bi,Buszello:2002uu,Choi:2012yg}, using the pioneering techniques of~\cite{Jacob:1959at} needs to be augmented by considering jet-inclusive final states.

While di-jet analyses exist and are already used to constrain the presence of BSM physics, analyses of electro-weak bosons in association with jets have received less attention (but do exist as well, see e.g.~\cite{Aad:2015ywd,CMS:2016qtb,Aaboud:2017nak,CMS:2017eej}). This is predominantly due to the fact that these channels are less common in established BSM scenarios and limits are typically dominated by cleaner $ZZ$ or $\gamma \gamma$ channels. However, particularly in the aforementioned scenarios of (partial) compositeness, these channels do provide important information about the couplings of a possible discovery. This motivates searches and a characterization program of jet-$\gamma$ and jet-$Z$ resonances (related by gauge invariance) as an important  probe of BSM physics at the LHC. Conversely, the lack of an observation in these channels would help to restrict the parameter space for such models. In either case, a detailed understanding of the possible dynamical scenarios is needed to perform an efficient analysis.

From a phenomenological perspective, jet-associated resonances are appealing as they combine potentially large signals, due to the presence of colored particles, with the precision of having a highly energetic photon or final state leptons, thereby being well-covered by existing trigger requirements. Searches so far have not given any hint for such resonances. 
 However, much more data is being collected and hopefully will soon be analyzed. We expect that studies of jet-$\gamma$ resonances will feature among the many channels in which searches for BSM physics are being carried out.

The purpose of this note is to present a model-independent leading-order analysis of the various possibilities of jet-$\gamma/Z$ interactions, aimed at extracting information about the spin of such resonances from the kinematic distribution of the outgoing photon or reconstructed $Z$ boson. (It is important to stress that EFT-based power-counting arguments might not be valid in case of a strongly interacting nature of such a resonance.) 

In Sec.~\ref{sec:hardinteractions}, we analyze the cases of spin $j=0, \half, 1, \threehalf, 2$, and construct the general form for the model-independent amplitude at parton level. In Sec.~\ref{sec:pheno}, we bridge these amplitudes to the hadron level for chosen benchmark scenarios and point out the sensitivity that can be expected from a dedicated spectroscopy program in these channels.

\section{Model-independent amplitudes}
\label{sec:hardinteractions}

In this section we present the amplitudes relevant for the study of jet-$\gamma$ and jet-$Z$ resonances. In the narrow width approximation, at leading order, we can decompose the amplitude into $2\to 1$ (production) and  $1\to 2$ (decay) on-shell processes, each characterized by a handful of coefficients coupling the different helicities. 

For the jet-$\gamma$ case, these coefficients are denoted by $a^{\pa\pa'}$ and $b^{\pa\pa'}$, where $\pa$ and $\pa'$ are the relevant ``partons'' and $b$ refers to the larger helicity component (below we will also label these coefficients with a subscript indicating the spin of the resonance). The P/CP properties are indicated  by putting a tilde on those coefficients related to amplitudes containing a $\gamma^5$ or a $\epsilon$ tensor. 

We retain the same notation for the  jet-$Z$ case, so that amplitudes with the same $a$ or $b$ coefficients reduce to the previous ones in the $m_Z\to 0$ limit. The additional coefficients arising from the longitudinal modes of the $Z$ are denoted by $c^{\pa\pa'}$ and the corresponding amplitudes vanish in the  $m_Z\to 0$ limit.

Since all $2\to1$ and $1\to 2$ amplitudes have the dimension of a mass, we divide by the appropriate power of the resonance mass $M$ to get the right overall dimension so that all coefficients $a, b, c$ are dimensionless. In the case where these coefficients arise from an effective field theory at the scale $\Lambda \gg M$, they will then scale by the appropriate powers of $M/\Lambda$ but we find it unnecessary to introduce an additional scale in the kinematics at this stage.

\subsection{The jet$+\gamma$ case}
\label{sec:jetphotonamp}
We begin by looking at $2\to 1$ processes involving 2 incoming massless particles of spin $1 \text{ or }\half$ creating a resonance of spin $0, \half, 1, \threehalf$ or $2$. These amplitudes are those of relevance to the production of the resonance but also, by using CPT, to the decay into a parton and a photon. In the next subsection we will consider the inclusion of the $Z$ boson.

For the process $\pa_1 \pa_2 \to X$ we introduce the partons' on-shell four-momenta, generically denoted by $p_1$ and $p_2$, ($p_1^2 = p_2^2 = 0$) as well as $p=p_1 + p_2$ and $q=p_1 - p_2$, obeying $p^2 = - q^2 = 2 p_1\cdot p_2= M^2$. The gluon and photon four-dimensional polarization vectors are transverse and the quark spinors obey the massless Dirac equation with the associated momenta.

As far as the polarizations of the resonance $X$ are concerned, we have $\not \! p U =M U$ and $\not \! p V = - M V$ for spin $\half$; $p^\mu S_\mu = 0$ for spin 1; $\not \! p U_\mu =M U_\mu$ and $\not \! p V_\mu = - M V_\mu$, $\gamma^\mu U_\mu = \gamma^\mu V_\mu = p^\mu U_\mu = p^\mu V_\mu = 0$ for spin $\threehalf$; $S_{\mu\nu} = S_{\nu\mu}$, $S_\mu^\mu = 0$ and $p^\mu S_{\mu\nu}=0$ for spin 2.

It is then straightforward to construct all combinations that are Lorentz invariant and obey the Ward identities. What is a bit more tedious is to eliminate all the linearly dependent combinations, particularly for the case containing the $\epsilon$-tensor where one needs to use Schouten's identity. This computation is greatly simplified by going to the center-of-mass frame (cms) of the resonance.

We always assume flavor conservation, so the quark polarizations $u, v$ refer to the same quark flavors. The color indexes of the resonance are denoted by $A$ (octet) or $a$ (triplet/anti-triplet), those of the partons by $B$, $B'$, $b$ or $b'$.  In the case of two gluons we denote by $\ve$ and $\ve'$ their polarizations and we can use either the totally anti-symmetric $f^{ABB'}$ or the totally symmetric $d^{ABB'}$ to respect the overall Bose symmetry, rendering selections rules a-la' Landau-Yang irrelevant in this case (a fact also mentioned in~\cite{Cacciari:2015ela}). The amplitudes read:
\vskip 1\baselineskip
\noindent Spin 0:
{\small
\begin{align}
   \gamma~g\to X:~~& \delta^{AB}\left(a^{\gamma g}_0 \left(M \vegamma_\mu {\veg}^\mu + 2 \vegamma_\mu q^\mu \veg_\nu q^\nu/M\right)
   +\tilde a^{\gamma g}_0 \vegamma_\mu \veg_\nu q_\lambda p_\rho \eps^{\mu\nu\lambda\rho}/M\right), \nn\\
   g~g\to X:~~& d^{ABB'} \left( a^{g g}_0 \left(M \ve_\mu {\ve'}^\mu + 2 \ve_\mu q^\mu \ve'_\nu q^\nu/M\right)+\tilde a^{g g}_0 \ve_\mu \ve'_\nu q_\lambda p_\rho \eps^{\mu\nu\lambda\rho}/M\right), \nn\\
   q~\bar q\to X:~~& T^{Ab}_{b'} \left( a^{q\bar q}_0 \bar v u + i \tilde a^{q\bar q}_0 \bar v \gfive u\right).\nn
\end{align}
}
Spin $\half$:
{\small
\begin{align}
     \gamma~q\to X:~~& \delta^a_b  \left(\frac{1}{\sqrt{2}} a^{\gamma q}_{1/2}( \ve_\mu \bar U \gamma^\mu u + 2 \ve_\mu q^\mu \bar U u/M)
      + \frac{1}{\sqrt{2}} \tilde a^{\gamma q}_{1/2}( \ve_\mu \bar U \gamma^\mu \gamma^5 u +2 \ve_\mu q^\mu \bar U \gamma^5 u/M)  \right),\nn\\
     \gamma~\bar q\to \bar X:~~& \delta^b_a \left(\frac{1}{\sqrt{2}} a^{\gamma \bar q}_{1/2}( \ve_\mu \bar v \gamma^\mu V - 2 \ve_\mu q^\mu \bar v  V/M)
      + \frac{1}{\sqrt{2}}\tilde a^{\gamma \bar q}_{1/2}( \ve_\mu \bar v \gamma^\mu \gamma^5 V + 2 \ve_\mu q^\mu \bar v \gamma^5 V/M)  \right),\nn\\
     g~q\to X:~~& T^{Ba}_b  \left( \frac{1}{\sqrt{2}} a^{g q}_{1/2}( \ve_\mu \bar U \gamma^\mu u + 2 \ve_\mu q^\mu \bar U u/M)
      + \frac{1}{\sqrt{2}} \tilde a^{g q}_{1/2}( \ve_\mu \bar U \gamma^\mu \gamma^5 u +2 \ve_\mu q^\mu \bar U \gamma^5 u/M) \right),\nn\\
     g~\bar q\to \bar X:~~& \tilde T^{Bb}_a  \left(\frac{1}{\sqrt{2}} a^{g \bar q}_{1/2}( \ve_\mu \bar v \gamma^\mu V - 2 \ve_\mu q^\mu \bar v  V/M)
      + \frac{1}{\sqrt{2}} \tilde a^{g \bar q}_{1/2}( \ve_\mu \bar v \gamma^\mu \gamma^5 V + 2 \ve_\mu q^\mu \bar v \gamma^5 V/M) \right).\nn
\end{align}
}
Spin 1:
{\small
\begin{align}
    \gamma~g\to X:~~& \delta^{AB}\left(a^{\gamma g}_{1}  \left(\vegamma_\mu {\veg}^\mu + 2 \vegamma_\mu q^\mu \veg_\nu q^\nu/M^2\right)
        S^*_\rho q^\rho + \tilde a^{\gamma g}_{1} S^*_\mu q^\mu \vegamma_\nu \veg_\rho q_\lambda p_\sigma
        \epsilon^{\nu\rho\lambda\sigma}/M^2 \right),\nn\\
    g~g\to X:~~& f^{ABB'}\left(a^{g g}_{1}\left(\ve_\mu {\ve'}^\mu + 2 \ve_\mu q^\mu \ve'_\nu q^\nu/M^2\right)S^*_\rho q^\rho +
      \tilde a^{g g}_{1} S^*_\mu q^\mu \ve_\nu \ve'_\rho q_\lambda p_\sigma
        \epsilon^{\nu\rho\lambda\sigma}/M^2\right),\nn\\
    q~\bar q\to X:~~& T^{Ab}_{b'}\left(a^{q \bar q}_{1} S_\mu^* q^\mu \bar v u/M + \frac{1}{\sqrt{2}} b^{q \bar q}_{1} S_\mu^* \bar v \gamma^\mu u
            + i \tilde a^{q \bar q}_{1} S_\mu^* q^\mu \bar v \gfive u /M + \frac{1}{\sqrt{2}} \tilde b^{q \bar q}_{1} S_\mu^* \bar v \gamma^\mu \gfive u\right).\nn
\end{align}
}
Spin $\threehalf$:
{\small
\begin{align}
     \gamma~q\to X:~~& \delta^a_b \left( \frac{\sqrt{3}}{2}
       a ^{\gamma q}_{3/2}( \ve_\mu \bar U_\nu \gamma^\mu u q^\nu/M + 2 \ve_\mu q^\mu \bar U_\nu u q^\nu/M^2)
       + b^{\gamma q}_{3/2}( \ve^\mu \bar U_\mu u - \half\ve_\mu \bar U_\nu \gamma^\mu u q^\nu/M) \right.\nn \\
      &+\left.  \frac{\sqrt{3}}{2}
      \tilde a^{\gamma q}_{3/2}( \ve_\mu \bar U_\nu \gamma^\mu \gamma^5 u q^\nu/M+ 2 \ve_\mu q^\mu \bar U_\nu \gamma^5 u q^\nu/M^2)
       +i \tilde b^{\gamma q}_{3/2}( \ve^\mu \bar U_\mu \gamma^5 u - \half\ve_\mu \bar U_\nu \gamma^\mu \gamma^5 u q^\nu/M)\right),\nn\\
     \gamma~\bar q\to \bar X:~~& \delta^b_a
      \left(\frac{\sqrt{3}}{2}
      a^{\gamma \bar q}_{3/2}( \ve_\mu \bar v \gamma^\mu V_\nu q^\nu /M - 2 \ve_\mu q^\mu \bar v  V_\nu q^\nu/M^2)
     +b^{\gamma \bar q}_{3/2}(\ve^\mu \bar v V_\mu  + \half\ve_\mu \bar v \gamma^\mu V_\nu q^\nu/M) \right. \nn\\
      &+\left. \frac{\sqrt{3}}{2}
      \tilde a^{\gamma \bar q}_{3/2}( \ve_\mu \bar v \gamma^\mu \gamma^5 V_\nu q^\nu/M + 2 \ve_\mu q^\mu \bar v \gamma^5 V_\nu q^\nu/M^2)
      +i \tilde b^{\gamma \bar q}_{3/2}(\ve^\mu \bar v \gamma^5 V_\mu  - \half \ve_\mu \bar v \gamma^\mu \gamma^5 V_\nu q^\nu /M)\right),\nn\\
     g~q\to X:~~& T^{Ba}_b  \left( \frac{\sqrt{3}}{2}
      a ^{\gamma q}_{3/2}( \ve_\mu \bar U_\nu \gamma^\mu u q^\nu/M + 2 \ve_\mu q^\mu \bar U_\nu u q^\nu/M^2)
      + b^{\gamma q}_{3/2}(\ve^\mu \bar U_\mu u - \half \ve_\mu \bar U_\nu \gamma^\mu u q^\nu/M) \right.\nn \\
      &+\left.\frac{\sqrt{3}}{2}
       \tilde a^{\gamma q}_{3/2}( \ve_\mu \bar U_\nu \gamma^\mu \gamma^5 u q^\nu/M + 2 \ve_\mu q^\mu \bar U_\nu \gamma^5 u q^\nu/M^2)
      +i \tilde b^{\gamma q}_{3/2}(\ve^\mu \bar U_\mu \gamma^5 u - \half\ve_\mu \bar U_\nu \gamma^\mu \gamma^5 u q^\nu/M)\right),\nn\\
     g~\bar q\to \bar X:~~& \tilde T^{Bb}_a \left(\frac{\sqrt{3}}{2}
       a^{g \bar q}_{3/2}( \ve_\mu \bar v \gamma^\mu V_\nu q^\nu/M - 2 \ve_\mu q^\mu \bar v  V_\nu q^\nu/M^2)
     + b^{g \bar q}_{3/2}(\ve^\mu \bar v V_\mu  + \half\ve_\mu \bar v \gamma^\mu V_\nu q^\nu/M)\right. \nn \\
      &+\left. \frac{\sqrt{3}}{2}
      \tilde a^{g \bar q}_{3/2}( \ve_\mu \bar v \gamma^\mu \gamma^5 V_\nu q^\nu /M+ 2 \ve_\mu q^\mu \bar v \gamma^5 V_\nu q^\nu/M^2)
      +i \tilde b^{g \bar q}_{3/2}(\ve^\mu \bar v \gamma^5 V_\mu  - \half \ve_\mu \bar v \gamma^\mu \gamma^5 V_\nu q^\nu/M)\right).\nn
\end{align}
}
Spin 2:
{\small
\begin{align}
   \gamma~g\to X:~~\delta^{AB}&\left(\sqrt{\frac{3}{2}} a^{\gamma g}_2 \left(\vegamma_\mu {\veg}^\mu + 2 \vegamma_\mu q^\mu \veg_\nu q^\nu/M^2\right)
        S^*_{\rho\lambda} q^\rho q^\lambda/M \right. \nn\\
 &+ \left. b^{\gamma g}_2 \left( M S^{\mu\nu *} \vegamma_\mu \veg_\nu + \veg_\mu q^\mu S^{\nu\rho*} \vegamma_\nu q_\rho/M
  + \vegamma_\mu q^\mu S^{\nu\rho*} \veg_\nu q_\rho/M - \half {\vegamma}^\mu \veg_\mu S^*_{\nu\rho} q^\nu q^\rho/M \right) \right. \nn \\
 &+ \sqrt{\frac{3}{2}}
 \tilde a^{\gamma g}_2 \vegamma_\mu \veg_\nu S^*_{\rho\lambda} q^\rho q^\lambda \epsilon^{\mu\nu\alpha\beta} q_\alpha p_\beta/M^3 \nn \\
 &+\left. \half \tilde b^{\gamma g}_2\left( \vegamma_\mu \veg_\nu S^*_{\gamma\rho} q^\rho \epsilon^{\mu\nu\gamma\lambda} q_\lambda /M+
  \vegamma_\mu {\veg}^\rho S^*_{\nu\rho} \epsilon^{\mu\nu\alpha\beta} q_\alpha p_\beta/M + \veg_\mu {\vegamma}^\rho S^*_{\nu\rho} \epsilon^{\mu\nu\alpha\beta} q_\alpha p_\beta/M \right)\right),\nn\\
   g~g\to X:~d^{ABB'}&\left(\sqrt{\frac{3}{2}} a^{g g}_2 \left(\ve_\mu {\ve'}^\mu + 2 \ve_\mu q^\mu \ve'_\nu q^\nu/M^2\right)
        S^*_{\rho\lambda} q^\rho q^\lambda /M\right. \label{Aspin2}\nn\\
 &+ \left. b^{g g}_2 \left( M S^{\mu\nu *} \ve_\mu \ve'_\nu + \ve'_\mu q^\mu S^{\nu\rho*} \ve_\nu q_\rho/M
  + \ve_\mu q^\mu S^{\nu\rho*} \ve'_\nu q_\rho/M - \half {\ve}^\mu \ve'_\mu S^*_{\nu\rho} q^\nu q^\rho/M \right) \right. \nn \\
 &+\left.\sqrt{\frac{3}{2}}
  \tilde a^{g g}_2 \ve_\mu \ve'_\nu S^*_{\rho\lambda} q^\rho q^\lambda \epsilon^{\mu\nu\alpha\beta} q_\alpha p_\beta/M^3\right) \nn \\
 + f^{ABB'} &\left( \half \tilde b^{g g}_2\left( \ve_\mu \ve'_\nu S^*_{\gamma\rho} q^\rho \epsilon^{\mu\nu\gamma\lambda} q_\lambda /M+
  \ve_\mu {\ve'}^\rho S^*_{\nu\rho} \epsilon^{\mu\nu\alpha\beta} q_\alpha p_\beta/M + \ve'_\mu {\ve}^\rho S^*_{\nu\rho} \epsilon^{\mu\nu\alpha\beta}
   q_\alpha p_\beta /M\right)\right),\nn\\
    q~\bar q\to X:~~T^{Ab}_{b'}&\bigg(\sqrt{\frac{3}{2}}
     a^{q \bar q}_2 S_{\mu\nu}^* q^\mu q^\nu \bar v u /M^2+ b^{q \bar q}_2 S_{\mu\nu}^* q^\nu \bar v \gamma^\mu u/M
    +\sqrt{\frac{3}{2}} i \tilde a^{q \bar q}_2 S_{\mu\nu}^* q^\mu q^\nu \bar v \gfive u/M^2 \nn \\&+ \tilde b^{q \bar q}_2 S_{\mu\nu}^* q^\nu \bar v \gamma^\mu \gfive u/M\bigg).\nn
\end{align}
}

The amplitudes involving two incoming gluons respect Bose symmetry under the exchange: $B \leftrightarrow B'$, $\ve \leftrightarrow \ve'$, $q \leftrightarrow -q$.

It is then straightforward to write down the non-zero production amplitudes $_{\mathrm{out}}\langle X, m| \pa, \lambda; \pa', \lambda'\rangle_{\mathrm{in}}$ for the process where parton $\pa$ with helicity $\lambda$ coming along the positive $\hat z$ axis ($\theta = 0$) and parton $\pa'$ with helicity $\lambda'$ coming along the negative $\hat z$ axis ($\theta = \pi$) create a resonance of spin $j$ and spin projection along $\hat z$ equal to $m = \lambda - \lambda'$. As already mentioned, the coefficients $a$ refer to the pairs of lowest helicity (RR or LL) while $b$ refers to RL or LR, if present.

The CPT theorem
\beq
_{\mathrm{out}}\langle \pa, \lambda;  \pa', \lambda' | X, m \rangle_{\mathrm{in}} = (-)^{j-m}~_{\mathrm{out}}\langle \bar X, -m| \bar \pa, -\lambda; \bar \pa', -\lambda'\rangle_{\mathrm{in}}, 
\eeq 
allows then one to write down the $1\to 2$ amplitudes for the decay of the resonance $X$ in its rest frame into one parton $\pa$ and a photon.
 
If one assumes the hermiticity of the effective interaction giving rise to the couplings (e.g. when the interaction arises by integrating out heavy mediators, as also discussed in~\cite{Panico:2016ary}), one can assume
$_{\mathrm{out}}\langle X, m| \pa, \lambda; \pa', \lambda'\rangle_{\mathrm{in}} = \; _{\mathrm{out}}\langle \pa, \lambda; \pa', \lambda' |X, m \rangle^*_{\mathrm{in}}$.
The above assumption, combined with the CPT theorem, 
forces the coefficients of the integer-spin amplitudes to be real and those of the half-integer ones to be conjugate of each other. In the simulations of Sec.~\ref{sec:pheno} we will always assume this to be the case.
 
\subsection{The jet+$Z$ boson case}
\label{sec:zbos}

For the decay into a jet and a $Z$ boson we need to generalize some of the amplitudes in the previous section to include a massive vector boson.
The notation is as before, but now $p_1^\mu$ is the momentum of the $Z$ boson, with $p_1^2 = m_Z^2$. The other parton is still massless and
$q=p_1-p_2$ and $p=p_1+p_2$ now obey  $q^2 = 2 m_Z^2 - M^2$, $p^2 = M^2$ and $p\cdot q=m_Z^2$. The energy of the massless parton in the center-of-mass frame is now $(M^2-m_Z^2)/2M$, so the on-shell normalization of the quark wave-function has changed. All remaining polarizations are the same and, for the $Z$, we have an additional longitudinal polarization. In order not to confuse the gluon and the $Z$ polarization, we refer to the gluon as $\varepsilon^\mu$ and denote the  $Z$ polarizations by $\zeta^\mu$.

We write down the production amplitudes for the process $Z\, \pa \to X$.
Of course, in this case one is really only interested in the conjugate process $X\to \,Z \pa$, but we stick with this notation for ease of comparison with the previous formulas. The coefficients are chosen in such a way that for $m_Z\to 0$ these amplitudes reduce to the previous one with a photon.
\vskip 1\baselineskip
\noindent Spin 0:
{\small
\begin{align}
   Z~g\to X:~~& \delta^{AB}\left(a^{Z g}_0 \left((M^2-m_Z^2)\zeta_\mu {\ve}^\mu + 2 \zeta_\mu q^\mu \ve_\nu q^\nu\right)/M
   +\tilde a^{Z g}_0 \zeta_\mu \ve_\nu q_\lambda p_\rho \eps^{\mu\nu\lambda\rho}/M\right).\nn
\end{align}
}
\noindent Spin $\half$:
{\small
\begin{align}
     Z~q\to X:~~& \delta^a_b  \Big(\frac{1}{\sqrt{2}}a^{Z q}_{1/2}( (M^2-m_Z^2)\zeta_\mu \bar U \gamma^\mu u + 2 M \zeta_\mu q^\mu \bar U u)/M^2 +
   2\, c^{Z q}_{1/2}\, m_Z^2 \zeta^\mu q_\mu  \bar U  u /M^3 \nn\\ &   +
    \frac{1}{\sqrt{2}} \tilde a^{Z q}_{1/2}( (M^2-m_Z^2) \zeta_\mu \bar U \gamma^\mu \gamma^5 u + 2 M \zeta_\mu q^\mu \bar U  \gamma^5 u)/M^2 +
     2 \, \tilde c^{Z q}_{1/2}\, m_Z^2 \zeta^\mu q_\mu  \bar U   \gamma^5 u / M^3  \Big) ,\nn\\ 
   Z~\bar q\to \bar X:~~& \delta^b_a \Big(\frac{1}{\sqrt{2}} a^{Z \bar q}_{1/2}( (M^2- m_Z^2)\zeta_\mu \bar v \gamma^\mu V - 2 M \zeta_\mu q^\mu \bar v  V)/M^2 +
      2\, c^{Z \bar q}_{1/2}\, m_Z^2 \zeta^\mu q_\mu  \bar v V /M^3 \nn\\ &   +
      \frac{1}{\sqrt{2}} \tilde a^{Z \bar q}_{1/2}( (M^2 - m_Z^2)\zeta_\mu \bar v \gamma^\mu \gamma^5  V + 2 M \zeta_\mu q^\mu \bar v \gamma^5  V)/M^2 +
      2 \, \tilde c^{Z \bar q}_{1/2}\, m_Z^2 \zeta^\mu q_\mu  \bar v \gamma^5  V /M^3 \nn
       \Big).
\end{align}
}
\noindent Spin 1:
{\small
\begin{align}
    Z~g\to X:~~\delta^{AB}&\bigg(a^{Z g}_{1}  \left((M^2-m_Z^2)\zeta_\mu {\ve}^\mu + 2 \zeta_\mu q^\mu \ve_\nu q^\nu\right)
        S^*_\rho q^\rho/M^2 \nn\\
      & +2\, c^{Z g}_{1}\,  m_Z^2\left( (M^2-m_Z^2) \ve^\mu S^*_\mu \zeta^\nu q_\nu+ \zeta^\mu q_\mu \ve^\nu q_\nu S^{*\rho} q_\rho
         \right)/M^4 \nn\\
      & + \tilde a^{Z g}_{1} S^*_\mu q^\mu \zeta_\nu \ve_\rho q_\lambda p_\sigma
        \epsilon^{\nu\rho\lambda\sigma}/M^2 + 2\, \tilde c^{Z g}_{1}\, m_Z^2 \zeta^\mu q_\mu \ve_\nu S^*_\rho  q_\lambda p_\sigma
        \epsilon^{\nu\rho\lambda\sigma}/M^4 \bigg).\nn
\end{align}
}
\noindent Spin $\threehalf$:
{\small
\begin{align}
     Z~q\to X:~~& \delta^a_b \Bigg( \frac{\sqrt{3}}{2}
       a ^{Z q}_{3/2}( (M^2 - m_Z^2)\zeta_\mu \bar U_\nu \gamma^\mu u q^\nu + 2 M \zeta_\mu q^\mu \bar U_\nu u q^\nu)/M^3\nn\\
       &+ b^{Z q}_{3/2}( (M^2 -m_Z^2)^2 \zeta^\mu \bar U_\mu u - \half M (M^2-m_Z^2)\zeta_\mu \bar U_\nu \gamma^\mu u q^\nu 
        + m_Z^2 \zeta^\mu q_\mu  q^\nu \bar U_\nu u )/M^4 \nn \\
        & + \sqrt 6\, c^{Z q}_{3/2}\, m_Z^2 \zeta^\mu q_\mu   q^\nu \bar U_\nu u/M^4 \nn\\
       & + \frac{\sqrt{3}}{2}
      \tilde a ^{Z q}_{3/2}( (M^2 - m_Z^2)\zeta_\mu \bar U_\nu \gamma^\mu \gamma^5 u q^\nu + 2 M \zeta_\mu q^\mu \bar U_\nu \gamma^5 u q^\nu)/M^3\nn\\
       &+i\, \tilde b^{Z q}_{3/2}\Big( (M^2 -m_Z^2)^2 \zeta^\mu \bar U_\mu  \gamma^5 u
            - \half M (M^2-m_Z^2)\zeta_\mu \bar U_\nu \gamma^\mu  \gamma^5 u q^\nu 
        + m_Z^2 \zeta^\mu q_\mu  q^\nu \bar U_\nu \gamma^5 u \Big)/M^4 \nn \\
        & +\sqrt 6 \, \tilde c^{Z q}_{3/2}\,  m_Z^2 \zeta^\mu q_\mu   q^\nu \bar U_\nu  \gamma^5 u/M^4\nn
        \Bigg),\\
     Z~\bar q\to \bar X:~~& \delta^b_a
      \Bigg(\frac{\sqrt{3}}{2}
      a^{Z \bar q}_{3/2}( (M^2 - m_Z^2) \zeta_\mu \bar v \gamma^\mu V_\nu q^\nu - 2 M\zeta_\mu q^\mu \bar v  V_\nu q^\nu)/M^3\nn\\
     &+b^{Z \bar q}_{3/2}((M^2 -m_Z^2)^2\zeta^\mu \bar v V_\mu  + \half M  (M^2-m_Z^2)\zeta_\mu \bar v \gamma^\mu V_\nu q^\nu
        + m_Z^2 \zeta^\mu q_\mu  \bar v V_\nu q^\nu)/M^4 \nn\\
      & + \sqrt 6\, c^{Z \bar q}_{3/2}\, m_Z^2 \zeta^\mu q_\mu  \bar v V_\nu q^\nu /M^4 \nn\\ 
      & +\frac{\sqrt{3}}{2}
      \tilde a^{Z \bar q}_{3/2}( (M^2 - m_Z^2) \zeta_\mu \bar v \gamma^\mu  \gamma^5 V_\nu q^\nu 
             + 2 M \zeta_\mu q^\mu \bar v   \gamma^5 V_\nu q^\nu)/M^3\nn\\
     &+i\, \tilde b^{Z \bar q}_{3/2}\Big((M^2 -m_Z^2)^2\zeta^\mu \bar v  \gamma^5 V_\mu  
         - \half M  (M^2-m_Z^2)\zeta_\mu \bar v \gamma^\mu \gamma^5 V_\nu q^\nu
        + m_Z^2 \zeta^\mu q_\mu  \bar v \gamma^5 V_\nu q^\nu\Big)/M^4 \nn\\
      & + \sqrt 6 \, \tilde c^{Z \bar q}_{3/2}\, m_Z^2 \zeta^\mu q_\mu  \bar v \gamma^5 V_\nu q^\nu /M^4 \Bigg).\nn
\end{align}
}
\noindent Spin 2:
{\small
\begin{align}
   Z~g\to X:~~\delta^{AB}&\Bigg(\sqrt{\frac{3}{2}} a^{Z g}_{2} \big((M^2 - m_Z^2)\zeta_\mu {\ve}^\mu + 2 \zeta_\mu q^\mu \ve_\nu q^\nu\big)
        S^*_{\rho\lambda} q^\rho q^\lambda /M^3 \nn\\
 & + b^{Z g}_{2} \bigg((M^2 - m_Z^2)^3 S^{\mu\nu *} \zeta_\mu \ve_\nu +(M^2 - m_Z^2)^2 \ve_\mu q^\mu S^{\nu\rho*} \zeta_\nu q_\rho  
 \nn\\ & ~~~~~~~~+ (M^4 - m_Z^4)\zeta_\mu q^\mu S^{\nu\rho*} \ve_\nu q_\rho  
   - \half M^2 (M^2 - m_Z^2){\zeta}^\mu \ve_\mu S^*_{\nu\rho} q^\nu q^\rho
 \nn\\ &~~~~~~~~ + m_Z^2{\zeta}^\mu q_\mu \ve^\nu q_\nu S^*_{\rho\lambda} q^\rho q^\lambda\bigg)/M^5 \nn \\
 &+2 \sqrt 2\, c^{Z g}_{2} \, m_Z^2\big((M^2-m_Z^2) S^{\mu\nu *} \ve_\mu q_\nu \zeta^\rho q_\rho + S^{\mu\nu *} q_\mu q_\nu \ve^\lambda q_\lambda  \zeta^\rho q_\rho  \big)/M^5\nn\\
 &+ \sqrt{\frac{3}{2}}
 \tilde a^{Z g}_{2} \zeta_\mu \ve_\nu S^*_{\rho\lambda} q^\rho q^\lambda \epsilon^{\mu\nu\alpha\beta} q_\alpha p_\beta /M^3 \nn \\
 &+ \tilde b^{Z g}_{2}
 \bigg( (M^2-m_Z^2)^2  \epsilon^{\mu\nu\gamma\lambda} S^*_{\mu\rho}\zeta^\rho \ve_\nu q_\gamma p_\lambda 
  -\half M^2 S^*_{\mu\nu}q^\mu q^\nu \epsilon^{\rho\lambda\alpha\beta} \zeta_\rho \ve_\lambda q_\alpha p_\beta
   \nn\\ & ~~~~~~~~-(M^2+m_Z^2) \zeta^\mu q_\mu \epsilon^{\rho\lambda\alpha\beta} \ve_\rho  S^*_{\lambda\sigma} q^\sigma q_\alpha p_\beta\bigg)/M^5
   \nn\\& +2\sqrt 2\, \tilde  c^{Z g}_{2}\, m_Z^2 \big(\zeta^\mu q_\mu  \epsilon^{\nu\rho\lambda\sigma} \ve_\nu 
     S^*_{\rho\gamma} q^\gamma q_\lambda p_\sigma \big)/M^5 \Bigg).\nn
\end{align}
}

Although the purpose of this work is to be as model-independent as possible, it is worth commenting on possible scenarios in which such couplings could arise. Historically, bosonic resonances of this type were considered in the context of Technicolor models~\cite{Ellis:1980hz} while fermionic resonances arose considering models of quark compositeness~\cite{Baur:1987ga}. These particles can be pair produced with ordinary QCD strength and this puts model independent bounds on the low mass region of the spectrum, typically below one TeV. The single production modes, via gluon-gluon or quark-gluon fusion, of interest to us have a higher mass reach but are more model-dependent.

As far as bosonic resonances are concerned, while the original motivation from technicolor  has greatly diminished due to the phenomenological difficulties of these models, recently there has been interest in the search for colored pseudo-Nambu-Goldstone bosons that arise in more recent models of partial compositeness~\cite{Cacciapaglia:2015eqa, Belyaev:2016ftv}. The production of such objects occurs via the anomaly and the cross section scales like $(K/f)^2$, where $K$ is the anomaly coefficient and $f$ the decay constant. For $K/f =1/\mbox{TeV}$ the production cross-section ranges from 0.1~pb for $M=500$~GeV to 0.1~fb for $M=3$~TeV at $\sqrt s= 13$~TeV. Here the bounds are still attractive for LHC searches, allowing, for some models, masses around one TeV for values of the decay constant of 1~TeV. 

As far fermionic resonances are concerned, the recent CMS search~\cite{CMS:2017eej} using the model~\cite{Bhattacharya:2009xg} sets a bound of 5.5~TeV on first and second family excited quarks for amplitude coefficient $a_{1/2}^{gq}$ $a_{1/2}^{\gamma q}$ of order $g_s$, $e$ respectively, and 1.8~TeV for excited $b$-quarks. The production cross section time branching ratio into jet-$\gamma$ are about 0.2~fb and 7~fb  at the upper limit of the excluded mass range. For amplitude coefficients reduced by a factor 10, the mass reach is below 2~TeV for light quarks while the search lacks sufficient sensitivity to set a bound for the $b$-quark.

\section{Elements of Hadron Collider Phenomenology}
\label{sec:pheno}
To gain a quantitative understanding of the phenomenology that we can expect at the LHC, we first consider the most motivated cases of scalars, spin half fermions and vector resonances. We later comment on the qualitative differences compared to the higher spin modes for representative examples, thus generalizing the analysis of~\cite{Aad:2015ywd,CMS:2016qtb,Aaboud:2017nak,CMS:2017eej} to all allowed coupling structures in light of CPT and Lorentz invariance.

We have implemented the couplings of the previous section into the {\sc{MadEvent}}~\cite{Alwall:2011uj,Alwall:2014hca} event generator with purpose-built {\sc{Helas}} routines~\cite{Murayama:1992gi}, and have checked these against implementations derived from the {\sc{FeynRules}}~\cite{Alloul:2013bka} and {\sc{Ufo}}~\cite{Degrande:2011ua} toolkits. As already mentioned, we only consider flavor symmetric cases and treat all quarks in the four-flavor scheme (and hence the involved pdfs) on an equal footing ($q=u,d,c,s$ in the following). We interface the generated parton-level events with Herwig++~\cite{Bahr:2008pv,Bellm:2015jjp} for showering and hadronization, and choose as benchmark $M=500~{\text{GeV}}$ for demonstration purposes. We will point out the influence of the mass scale $M$ on our analysis below, where we will also comment on detector resolution effects. Throughout, we focus on 13 TeV collisions.

Due to the very character of the above amplitudes being decompositions of physical scattering amplitudes in the narrow width approximation, we choose a small reference width $\Gamma_X/M=10^{-4}$. In this work we will not discuss the extraction of the coupling sizes comprehensively, but we note that our analysis of the signal events is entirely insensitive to the exact choice as long as $\Gamma_X/M\ll 1$. Since the amplitudes of Sec.~\ref{sec:hardinteractions} are not valid in the off-shell regime of any involved legs, we can expect measurements of the CP character of the resonance along the lines 
of~\cite{Plehn:2001nj,Hankele:2006ma,Klamke:2007cu,Hagiwara:2009wt,Englert:2012ct,Englert:2012xt,Dolan:2014upa}, when relevant, to be significantly limited already at this stage. The non-validity of the above amplitudes for off-shell momenta also does not allow us to perform multi-jet matching for the signal.

We have simulated the contributing backgrounds in an identical way and specifically focus on QCD-induced $\gamma/Z$+jet production which are the largest contributing backgrounds in the SM. The LHC experiments typically estimate these with data-driven methods (see e.g.~\cite{Aad:2015ywd}) and this part of our analysis solely serves as numerical guide to highlight the potential sensitivity for our benchmark. To allow us to compare signal and background on an equal footing, we do not include jet-matching effects for the backgrounds.

Since the resonances we study in this work are motivated from general amplitude Lorentz structures, and since these carry colour charges, such a discovery can be established not only $\gamma/Z+\hbox{jet}$ production, but also in multi-jet final states. Analyses of the latter have been carried out in two jet~\cite{Sirunyan:2016iap,Aaboud:2017yvp}, as well as in four jet final states~\cite{ATLAS:2012ds,Khachatryan:2014lpa}. A discovery will crucially depend on the sizes of the coefficients quoted in Secs.~\ref{sec:jetphotonamp} and~\ref{sec:zbos}, as these can be chosen to avoid discovery in the multi-jet channels.

The LHC already performs bump hunts for resonances in the $\gamma+\hbox{jet}$ channel as detailed in~\cite{Aad:2015ywd,CMS:2016qtb,Aaboud:2017nak,CMS:2017eej}. These are performed by a data-driven estimate of the sensitive invariant mass distribution with the aim to reveal excesses in model-independent approach. If such a search is successful, the question of the precise coupling structure of the new discovery arises. The $Z$ boson takes a special role in this case due to gauge invariance (however at a much smaller rate due to leptonic branching ratios which deliver the cleanest signatures for subsequent analyses). The imminent spectroscopy programme after such a discovery will then need to be informed by a range of searches, in particular because the exclusive rate of decays into the final states we discuss in this work will be influenced by the coefficients of the di-jet channels. For the purpose of this work we will assume a discovery in the jet+$\gamma$ channel at 300/fb (this might be accompanied by a similar observation in the multi-jet channels) and we outline a spectroscopy follow-up programme in the jet+$\gamma/Z$ channels. We will come back to the importance of multi-jet final states at the end of this section.

Different spin expectations can be discriminated by characteristic angular distributions~\cite{Cabibbo:1965zzb,Trueman:1978kh,Collins:1977iv,DeRujula:1983ak,DellAquila:1985mtb,DellAquila:1985jin,Nelson:1986ki} 
\begin{figure*}[!t]
\parbox{0.46\textwidth}{
    \centering
	\includegraphics[width=8cm]{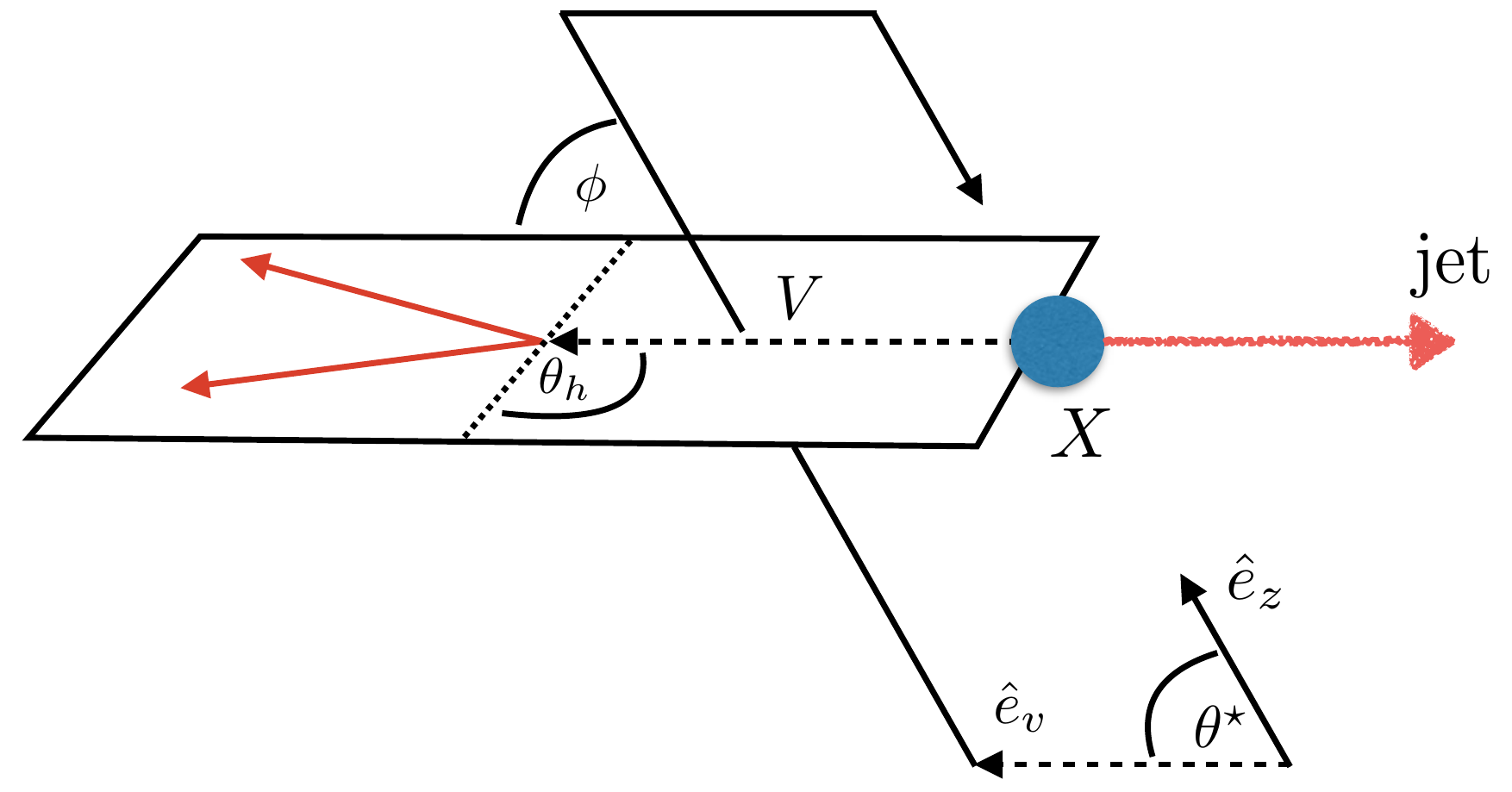}
}
\hfill
\parbox{0.46\textwidth}{\caption{\label{fig:angles} Illustration of the angles sensitive to the spin and polarization information discussed in this work. $\hat{e}_z$ denotes the normalised direction of the proton-proton beam axis, $\hat{e}_v$ denotes the direction of $Z$ or $\gamma$ boson in the particle $X$ rest frame. In case of $Z+\hbox{jet}$ production the $Z$ boson lepton decay angles include important information: $\theta_h$ denotes the angle of the negatively charged lepton against the resonance $X$ in the $Z$ boson rest frame. $\phi$ denotes the angle between the production and decay planes.
}}
\end{figure*}
(see also \cite{Buszello:2002uu,Gao:2010qx,Englert:2010ud}). The angles that are sensitive to the spin structure of the interactions which are relevant to our final states are (Fig.~\ref{fig:angles})
\begin{alignat}{5}
\label{eq:costhetahel}
\cos\theta_h &=   { {\vec{p}}_{\ell^-} \cdot {\vec{p}}_{X}  \over  \sqrt{  {\vec{p}}^{\,2}_{\ell^-}\, {\vec{p}}^{\,2}_{X}    }} \bigg|_{Z} ,\\
\label{eq:costh}
\cos\phi &=  { (\hat e_z\times \hat e_{v})\cdot ({\vec{p}}_{\ell^-} \times {\vec{p}}_{\ell^+}) \over 
 \sqrt{( {\vec{p}}_{\ell^-} \times {\vec{p}}_{\ell^+} )^2} } \bigg|_X \,,\\
 \label{eq:costhetastar}
 \cos\theta^\ast &=   { {\vec{p}}_{V} \cdot {\hat e_{z}}  \over \sqrt{  {\vec{p}}^{\,2}_{V} }} \bigg|_{X}\,.
\end{alignat}
The subscripts $X,Z$ refer to the rest frames in which these angles are defined. The momenta are defined from the decay products, i.e. 
\begin{equation}
	\vec{p}_X=\vec{p}_{\gamma/Z} + \vec{p}_j
\end{equation}
with 
\begin{equation}
	\vec{p}_Z=\vec{p}_{\ell^-} + \vec{p}_{\ell^+}
\end{equation}
in case of $\hbox{jet}+Z$ production.
Note that the helicity angle $\theta_h$ and the azimuthal angle $\phi$ are not observable for decays $X\to \gamma j$, hence limiting the available range of sensitive observables to $\theta^\ast$. We have introduced $\phi$ for completeness but we find that it contains no discriminative power for the scenarios we study in this work, and we will not further consider this angle in the following.

The discriminating power of these angles (we will discuss the contributing
backgrounds further below) lies in the fact that the boosts into
respective rest frames remove some kinematic dependence on the final
states and the mass of $X$ in particular. For the jet-$\gamma$ case, in the lab frame, the scattering is fully described through a combination of transverse momentum $p_T$ and pseudorapidity $\eta$ of the photon.

\begin{figure*}[!t]
\centering
 \includegraphics[height=6cm]{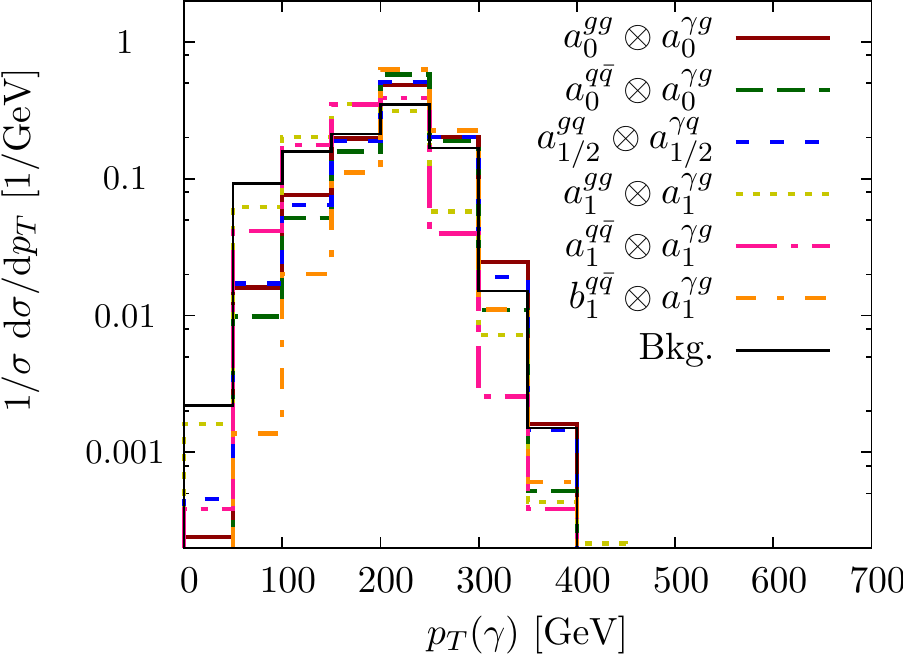}\hfill
 \includegraphics[height=6cm]{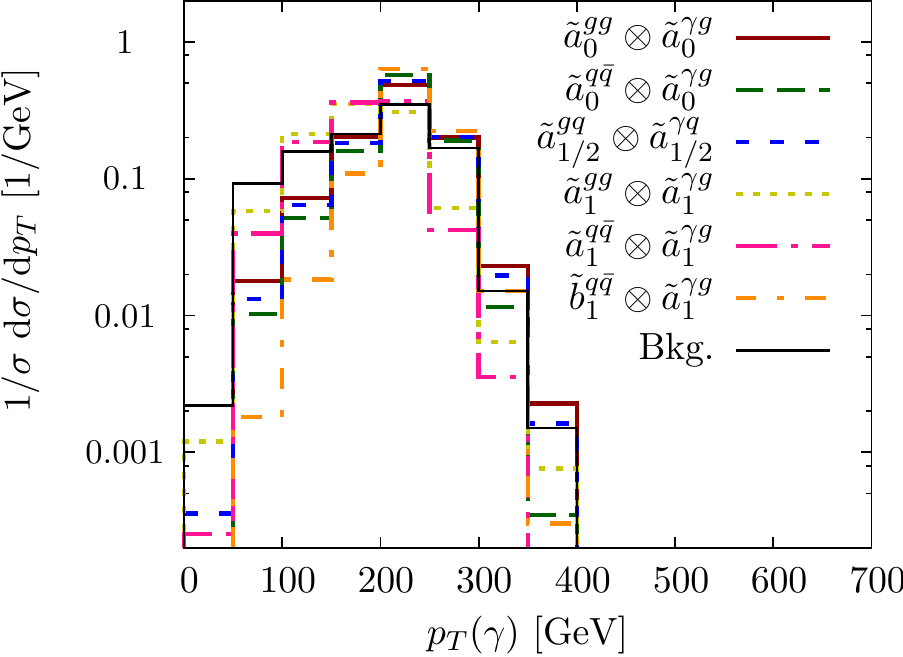}\\
  \caption{\label{fig:jetphoton} Normalized transverse momentum distribution for jet+$\gamma$ production, focussing on the different spin $j$  (here specifically $j\leq 1$) and types of coupling detailed in Sec.~\ref{sec:hardinteractions}. The lower index denotes the spin $j$. The symbol $\otimes$ separates the coefficients involved in the production and decay of the resonance. The resonance mass $M$ is set at 500~GeV, merging or detector effects are not included here.}
\end{figure*}

While removing this energy
dependence has the benefit of projecting out the helicity
decomposition of the interactions of Sec.~\ref{sec:hardinteractions} (sculpted to 
some
extent by finite detector coverage), the distribution of the signal
events according to energy-sensitive observables such as the
transverse photon momentum can provide evidence of the dominant
production mechanism via pdf effects. Another avenue we will discuss
in the following is 
the prospects of quark-gluon tagging \cite{ATLAS:2016wzt, CMS:2017wyc,Gallicchio:2012ez}, which can in
principle further discriminate the $X$ decay phenomenology, thus
providing important information in discriminating the different spin
hypotheses. This strategy is particularly motivated, as the threshold induced by the expected large mass of $X$ helps to choose working points that are particularly attractive to quark-gluon tagging~\cite{FerreiradeLima:2016gcz}.

Our event selection is performed fairly inclusively
only reflecting the basic trigger thresholds for the signal
events to be recorded. Representative generator level ccross sections for the coupling combinations that we study in this paper are tabulated in Tabs.~\ref{tab:axsecs} and~\ref{tab:zxsecs}.

\begin{table}[!t]
\begin{center}
\begin{tabular}{c c c c}
\hline
scenario & $\sigma(M=500~\text{GeV})$ [pb] & $\Gamma_X$ [GeV] & BR$(X\to \gamma j)$ \\
\hline
${a}^{gg}_0 \otimes {a}^{\gamma g}_0$ & 83.58 & 0.36 & 0.55 \\
${a}^{q\bar q}_0\otimes {a}^{\gamma g}_0$ & 48.83& 0.60 & 0.33 \\
$a^{gq}_{1/2} \otimes {a}^{\gamma q}_{1/ 2}$  & 82.54 & 0.93 & 0.43\\
$ a^{gg}_1 \otimes {a}^{\gamma g}_{1}$  &  127.2  & 0.16 & 0.40 \\
$a^{q \bar{q} }_{1} \otimes {a}^{\gamma g}_{1}$ & 22.22 &  0.20 & 0.33 \\
$b^{q \bar{q} }_{1} \otimes {a}^{\gamma g}_{1}$ & 25.45 & 0.20 & 0.33 \\
$a^{gq}_{3/2} \otimes a^{\gamma q}_{3/2} $ & 214.9  & 0.46 & 0.43\\
$a^{gg}_{2} \otimes a^{\gamma g}_2 $   &  68.71 & 0.07  & 0.55\\
\hline
\end{tabular}
\caption{
Central value cross sections for the $\gamma$+jet scenarios discussed in Sec.~\ref{sec:pheno} for coupling values of $10^{-1}$. The results for P/CP-violating parameters are identical.\label{tab:axsecs}}
\end{center}
\end{table}

\begin{table}[!t]
\begin{center}
\begin{tabular}{c c c c}
\hline
scenario & $\sigma(M=500~\text{GeV})$ & $\Gamma_X$ [GeV] & BR$(X\to Z j) $ \\
\hline
${a}^{gg}_0 \otimes {a}^{Z g}_0$ & 66.00 & 0.36 & 0.54  \\
$a^{gq}_{1/2} \otimes {a}^{Z q}_{1/ 2}$  & 314.0 & 0.92  & 0.43 \\
$ a^{gg}_1 \otimes {a}^{Z g}_{1}$  &  105.9  & 0.17 & 0.40 \\
$ a^{gg}_1 \otimes {c}^{Z g}_{1}$  &  30.70  & 0.11 & 0.08  \\
$a^{gq}_{3/2} \otimes a^{Z q}_{3/2} $ &  74.03& 0.46 & 0.42 \\
$a^{gg}_{2} \otimes a^{Z g}_2 $   &  53.97 &  0.07 &  0.55\\
\hline
\end{tabular}
\caption{Central value cross sections for the $Z$+jet scenarios discussed in Sec.~\ref{sec:pheno} for coupling values of $10^{-1}$. The results for P/CP-violating parameters are identical.\label{tab:zxsecs}}
\end{center}
\end{table}

\begin{figure*}[!t]
\centering
 \includegraphics[height=6.cm]{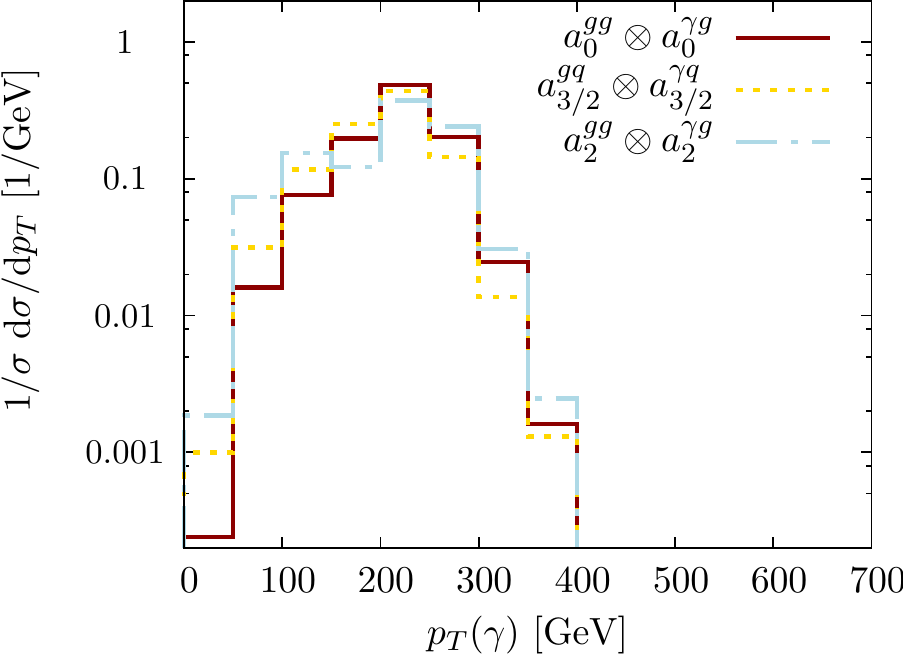}\hfill
 \includegraphics[height=6.cm]{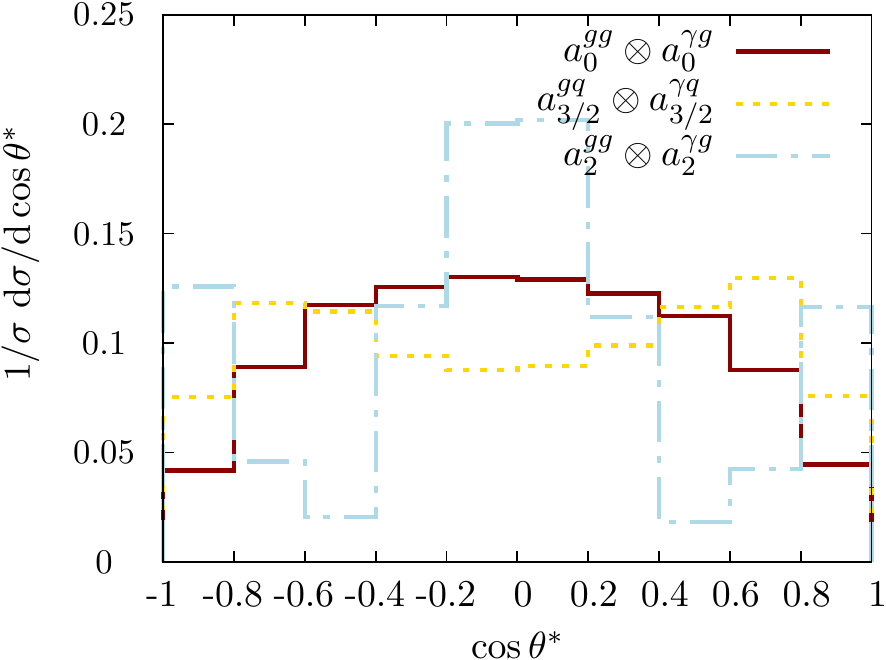}\\
  \caption{\label{fig:jetphoton3} Normalized transverse momentum and $\cos\theta^\ast$ distributions for jet+$\gamma$ production, focussing on representative spin $3/2$ and spin $2$ couplings of Sec.~\ref{sec:hardinteractions}. (Spin $0$ included for comparison.) See caption of Fig.~\ref{fig:jetphoton} for further details. No merging or detector resolution effects are included.}
\end{figure*}

For the case of photon-associated jet production we require an
isolated photon (defined as isolated if the hadronic energy deposit in
an area $R<0.3$ around the photon candidate is less than 5\% of the photon's
transverse momentum) with
\begin{equation}
  p_{T,\gamma} >30~\text{GeV},~\hbox{and}~|\eta_\gamma| < 2.33\,,
\end{equation}
to guarantee the event to be triggered~\cite{ATL-COM-DAQ-2017-117}. In case of final
state leptons, we require two isolated leptons (hadronic energy
deposit less than 10\% of the lepton candidate's transverse momentum
in $R<0.3$) with opposite charge and
\begin{equation}
  p_{T,\ell} >30~\text{GeV},~\hbox{and}~|\eta_\ell| < 2.5\, .
\end{equation}
Again these criteria reflect the standard trigger thresholds~\cite{ATL-COM-DAQ-2017-117}. On top of these thresholds we require 
\begin{wrapfigure}[26]{r}{0.62\textwidth}	
\vspace{-0.6cm}
\hfill\parbox{0.62\textwidth}{
    \centering
 \includegraphics[height=6cm]{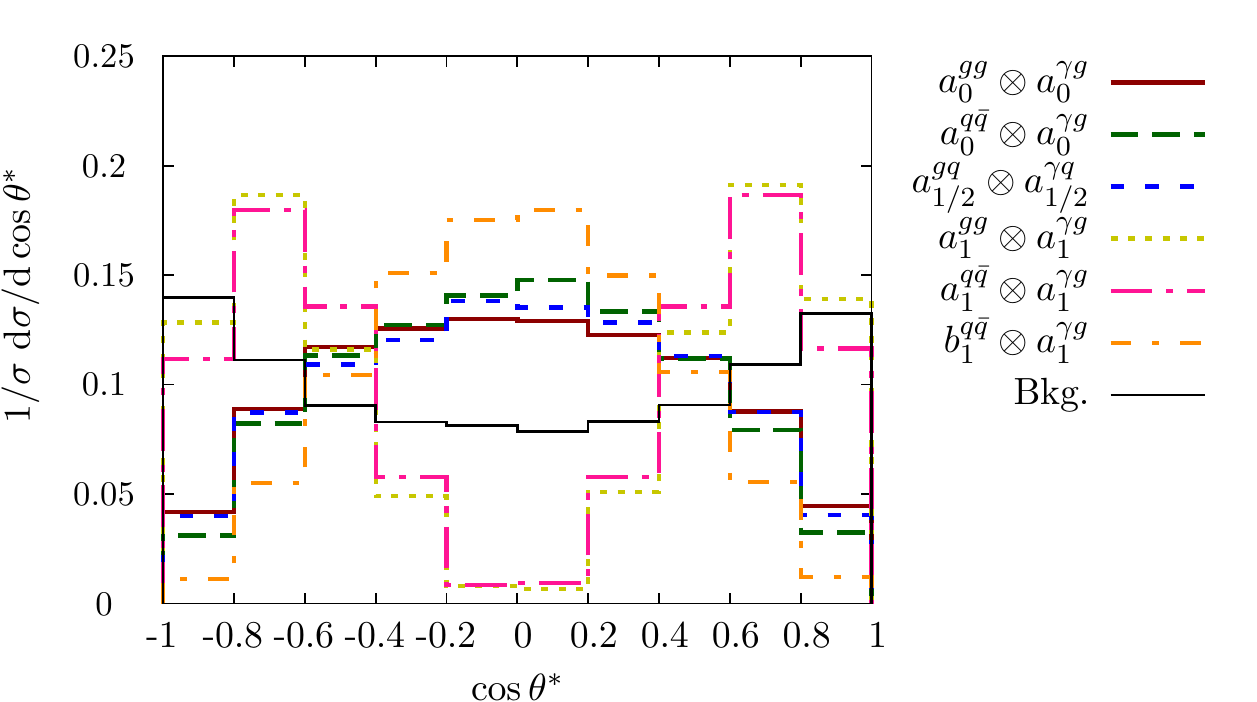}\\[0.3cm]
 \includegraphics[height=6cm]{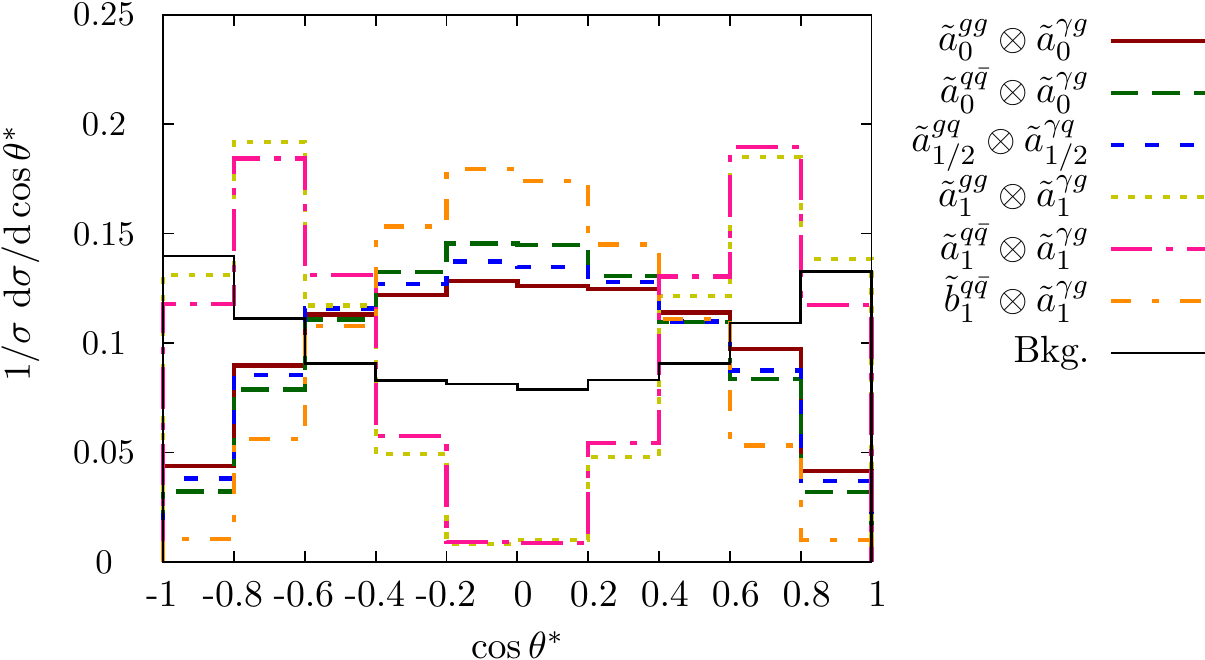}
\caption{\label{fig:jetphoton2} Normalized $\cos\theta^\ast$ distribution for jet+$\gamma$ production for various spin and couplings (not considering detector resolution or jet-merging effects). The distributions do not allow to discriminate between the P/CP properties of the interactions.}
}
\end{wrapfigure} 
the leptons to be compatible with the $Z$ pole mass
\begin{equation}
|m_{\ell^+\ell^-}-m_Z|<5~\text{GeV}\,,
\end{equation}
in the leptonic final state case.

The jets are clustered with the anti-kT algorithm~\cite{Cacciari:2008gp} with resolution
parameter $D=0.4$ using {\sc{FastJet}}~\cite{Cacciari:2011ma} and we define jets from the thresholds
\begin{equation}
  p_{T,j} >30~\text{GeV},~\hbox{and}~|\eta_j| < 4.5\,.
\end{equation}
We require the leading jet in $p_T$ to be inside $|\eta_{j_1}| < 2.33$ in a back-to-back configuration with the reconstructed photon or $Z$ boson ($p_Z=p_{\ell^-} +p_{\ell^+}$) in the azimuthal angle--pseudo-rapidity
plane $R(j,Z/\gamma)>2.5$. Finally we require consistency of the reconstructed resonance with our mass hypothesis within 50~GeV
\begin{equation}
\label{eq:massselection}
	|m_{j\gamma/Z}-M| < 25~\text{GeV}\,.
\end{equation}
This latter criterion, while not relevant for our signal distributions, is crucial for a comparison of the signal with the expected background. For comparison, ATLAS searches are sensitive to width/mass ratios of 2\% in Ref.~\cite{Aad:2015ywd}, which is well-covered by our representative invariant mass window cut in our signal-like selection, where our approximations can be trusted.

For the jet$+\gamma$ case, where we essentially only have a single angle at our disposal to discriminate the various hypotheses, we can already identify the qualitative overall behavior of the final state. The $p_T$ distribution of the photon (Figs.~\ref{fig:jetphoton} and~\ref{fig:jetphoton3}), while giving some indication of the dominant partonic subprocess through the parton distribution functions as well as spin and coupling character, is largely dominated by the threshold of the particle $X$, whose mass gets equally distributed into transverse momentum for central production. The differen-

tial measurement of $\theta^\ast$ (Figs.~\ref{fig:jetphoton3} and~\ref{fig:jetphoton2}), on the other hand, allows us to formidably discriminate between different spin hypotheses. Some of the remaining qualitative degeneracies can be lifted (see below).\footnote{Note that the acceptance selections (which are determined by trigger thresholds and mostly finite detector coverage) bias the constructed distribution, see~\cite{Englert:2010ud} for a detailed discussion in the context of QCD singlets.}

The background distribution is fundamentally different from the signal distribution. As alluded
\begin{figure*}[!t]
 \vspace{-0.6cm}
 \parbox{0.48\textwidth}{\includegraphics[height=6cm]{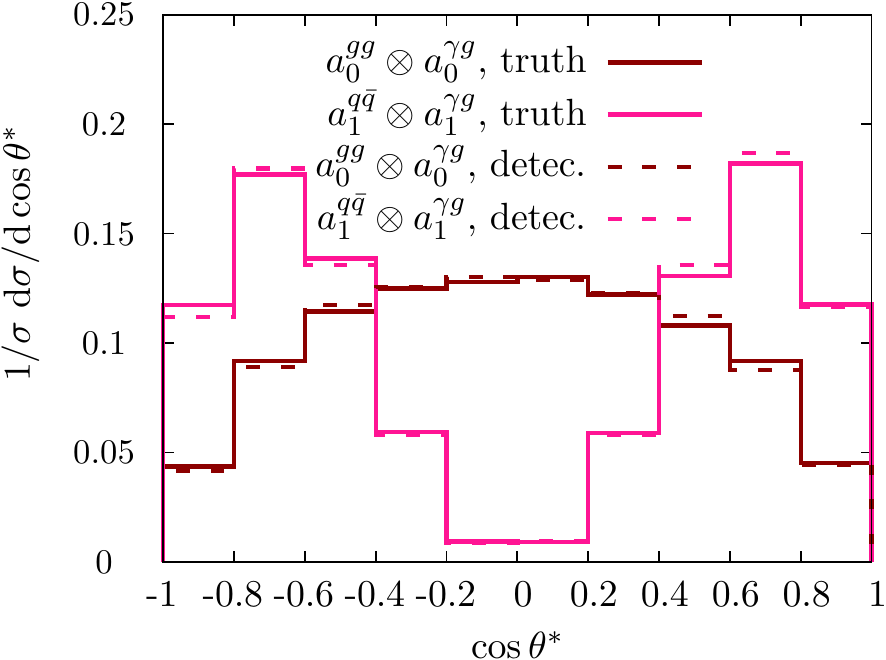}
\caption{\label{fig:exresol} Impact of energy mis-measurements on the $\cos\theta^\star$ distribution for our benchmark scenario of $M=500$~GeV for two representative spin and coupling hypotheses.}}
\hfill
\parbox{0.48\textwidth}{
\vspace{0.6cm}
 \includegraphics[height=6cm]{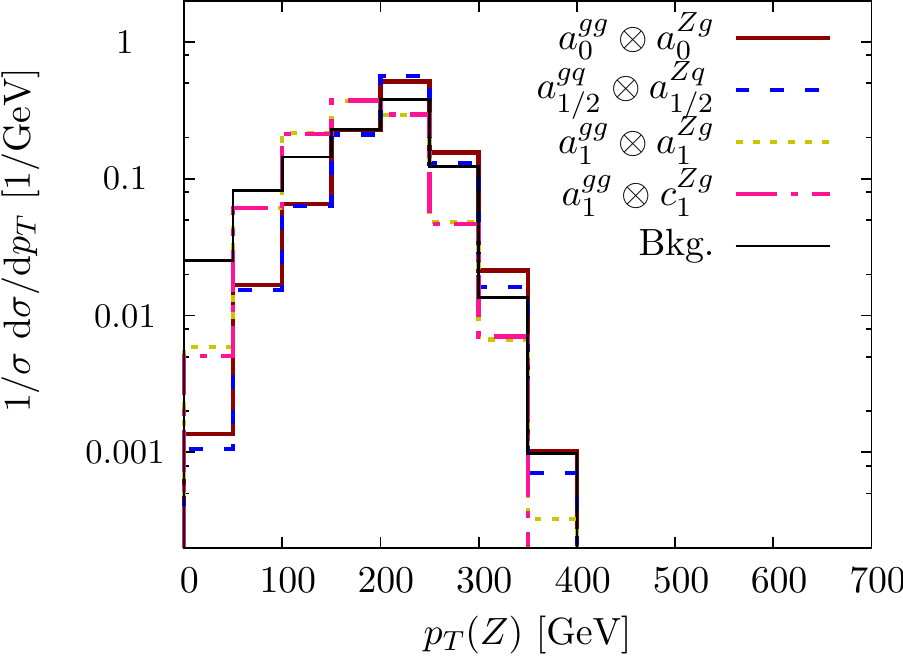}
\caption{\label{fig:jetz} Normalized transverse momentum distribution for jet+$Z$ production, focussing on representative spin $0,1/2,1$ and coupling properties of Sec.~\ref{sec:hardinteractions} (no jet merging and detector resolution effects). See caption of Fig.~\ref{fig:jetphoton} for further details on the notation.}
}
\end{figure*}
to above, none of the observables of the hard $2\to 2$ scattering reflects the P/CP character of the couplings and more involved processes that access off-shell information need to be considered (see Fig.~\ref{fig:jetphoton2}).

\begin{figure*}[!t]
    \centering
\parbox{0.65\textwidth}{
 \includegraphics[height=6cm]{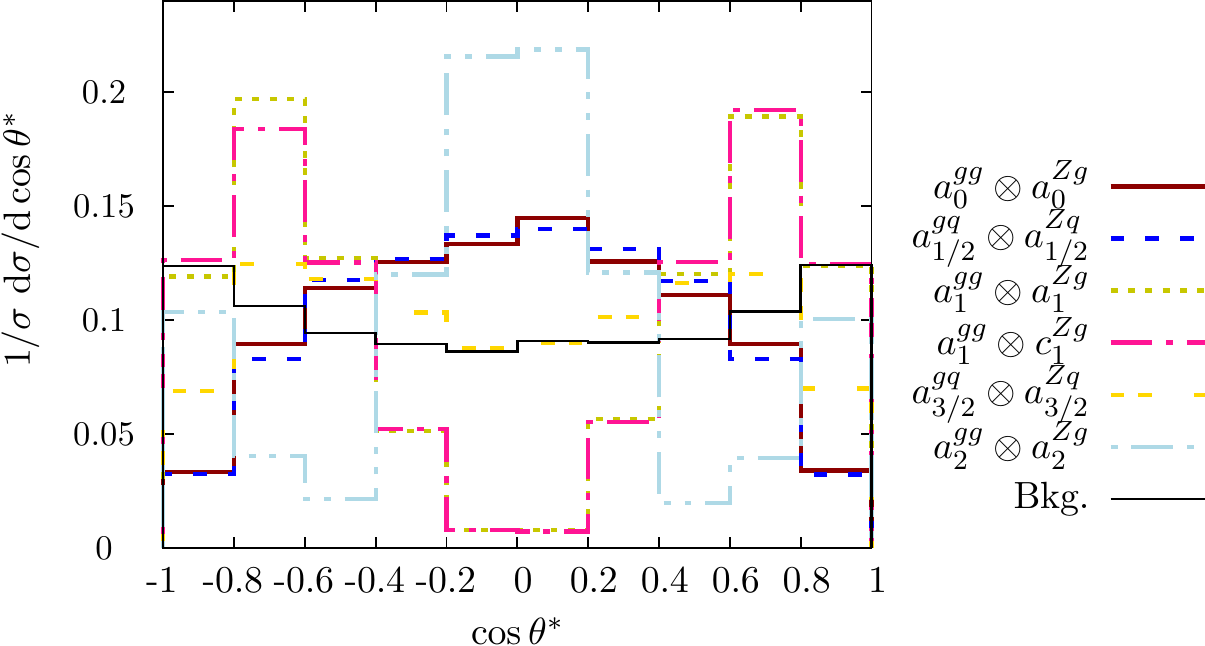}\\[0.3cm]
 \includegraphics[height=6cm]{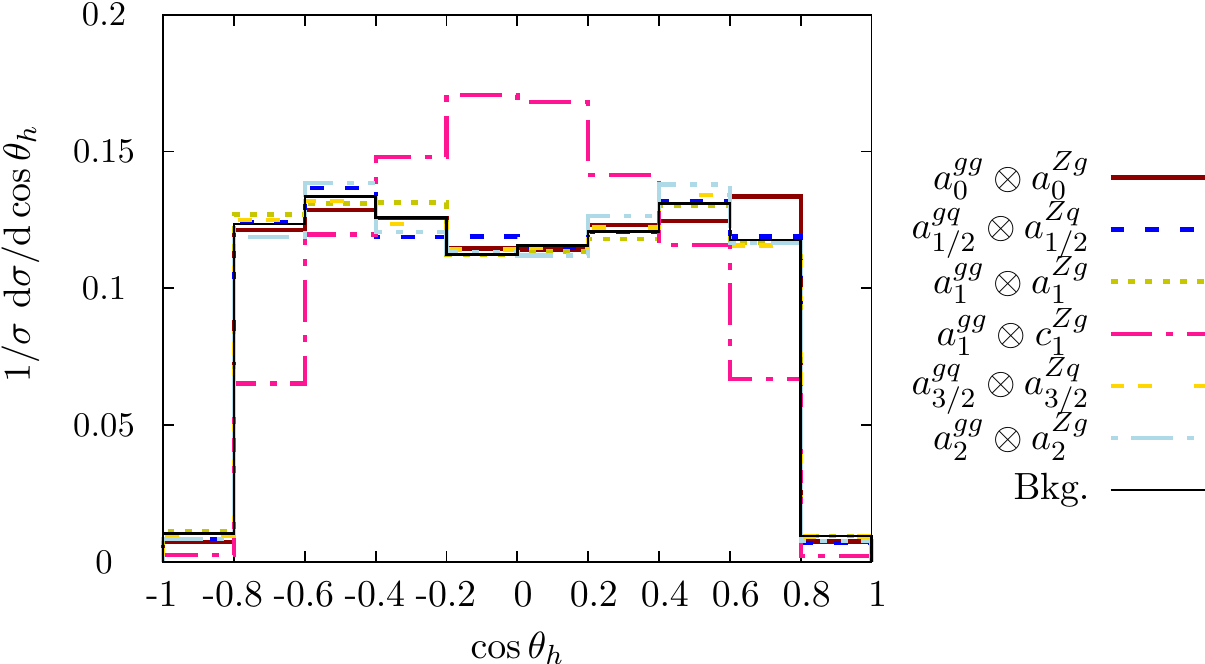}}
\hfill  \parbox{0.3\textwidth}{
\vspace{7cm}
\caption{\label{fig:jetz2} Normalized $\cos\theta^\ast$ and helicity angle distributions for jet+$Z$ production for the different spin and coupling properties as detailed in Sec.~\ref{sec:hardinteractions} (no jet merging and detector resolution effects are included). We include the expected background distributions. See caption of Fig.~\ref{fig:jetphoton} for further details on the notation.}}
\end{figure*}

All results in Figs.~\ref{fig:jetphoton} -- \ref{fig:jetphoton2} do not include the expected effects of energy mis-measurements. As the sensitivity in these angle depends on the correct reconstruction of the $X$ rest frame, detector resolution effects can in principle limit the sensitivity. In practice, when the mass scales are large, the expected calibration of photons and jets has shown to be under extremely good control and is likely to improve even further in the future~\cite{Tapper:2013yva}. However, to provide a more quantitative understanding of how detector effects modify the angular distributions we show a comparison of $\cos\theta^\star$ including detector effects in Fig.~\ref{fig:exresol} (we have adopted the energy parameterisation of {\sc Delphes3} \cite{deFavereau:2013fsa}).

This phenomenological situation remains largely unchanged when considering the $Z$ boson case (Figs.~\ref{fig:jetz} and ~\ref{fig:jetz2}) as far as the $p_T$ and $\eta$ distributions are concerned and we choose not to show these distributions for this reason. However, $\cos\theta^\ast$ remains an important additional handle to constrain and discriminate different scenarios (albeit at a lower rate due to the leptonic decays of the $Z$ boson that we consider here). The distribution of $\cos\theta_h$ essentially discriminates the decays of the transversely polarized $Z$ (coefficients $a$ and $b$) from the longitudinal one (coefficient $c$). See some examples in Fig.~\ref{fig:jetz2}.

It is important to add that the background distribution is highly sculpted towards the bulk of signal hypotheses, making further discrimination beyond the aforementioned cases increasingly difficult.

\begin{figure}[!t]
\centering
 \subfigure[\label{fig:taga}\it CLs discrimination based on $\cos\theta^\ast$ before quark/gluon tagging.]{\includegraphics[width=8cm]{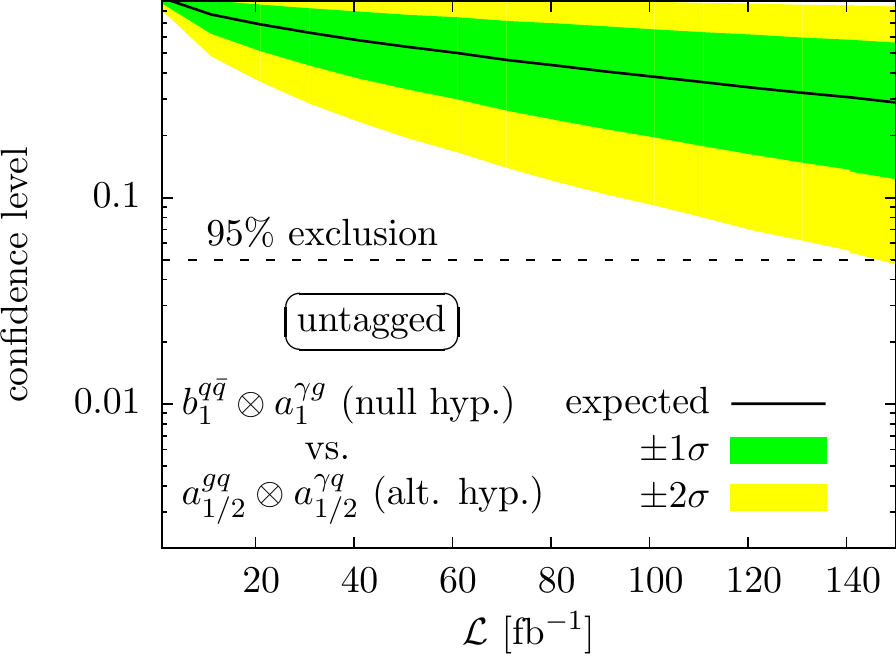}}\hfill
 \subfigure[\label{fig:tagb}\it CLs discrimination based on $\cos\theta^\ast$ including quark/gluon tagging.]{\includegraphics[width=8cm]{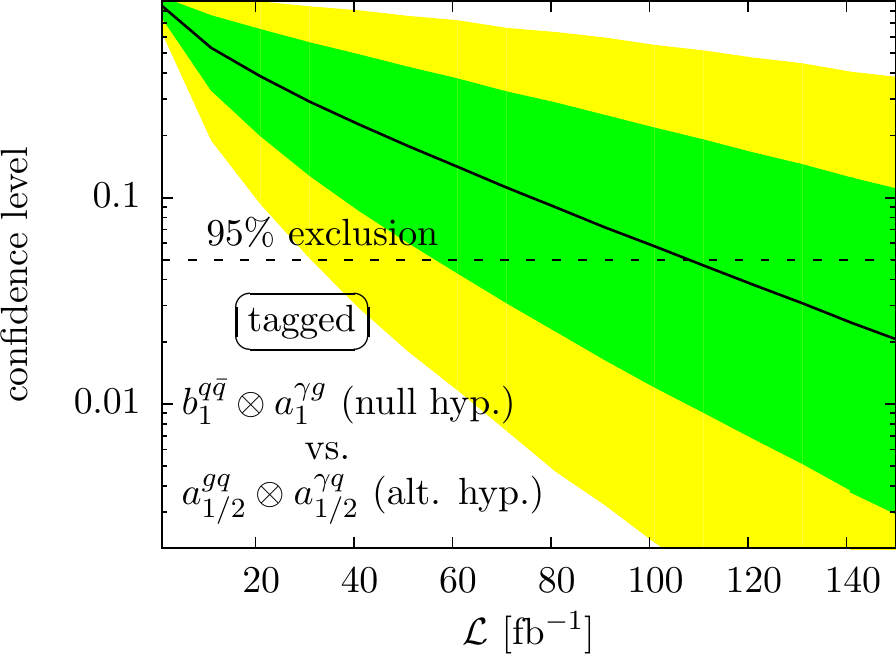}} \caption{\label{fig:jettagging} Importance of quark/gluon tagging as a function of integrated luminosity. The overall signal normalization is aligned with Fig.~\ref{fig:jetaextr}, where we assume a $S/\sqrt{B}=5$ for 300~$\text{fb}^{-1}$ luminosity in the untagged category. This amounts to a signal cross section after cuts of $23.4~\text{fb}$.
 }
 \end{figure}

An additional handle for disentangling the spin hypothesis is through discriminating the production modes. While the transverse momentum distributions of the $Z$ boson or the photon give some understanding of the dominant perturbative partonic subprocess, another avenue to further access this information is via identifying, at least approximately~(see e.g. the recent Ref.~\cite{Gras:2017jty}), the quark- or gluon-like character of the leading final state jet. This analysis step, which is entirely complementary to the analysis of angles $\theta_h$ and/or $\theta^\ast$, needs to be understood as an additional criterion that is invoked on  a final selection that separates signal from background. The drop in signal rate  
before and after quark/gluon tagging is applied will provide additional power separating integer spin from the ${1\over 2}$ and ${3\over 2}$ hypotheses. This is shown representatively in Fig.~\ref{fig:jettagging} for the concrete example of separating the $b^{q\bar q}_{1}\otimes a^{\gamma g}_1$ coupling combination as null-hypothesis  from the $a^{gq}_{1/2}\otimes a^{\gamma q}_{1/2}$ (alternative) hypothesis. The exclusion limits of Fig.~\ref{fig:jettagging} are calculated using the CLs method~\cite{Junk:1999kv,Read:2000ru,Read:2002hq} for a binned log-likelihood based on $\cos\theta^\ast$ observable in the jet+$\gamma$ channel of~Fig.~\ref{fig:jetphoton2}, assuming ``even''  coupling structures. 

This particular choice for a representative example is motivated from the overall similarities in the $\cos\theta^\ast$ observable, however, with a clear separation of quark- vs. gluon-initiated hard jet. As representative working point for quark-tagging and gluon-rejection, we use efficiencies $(\epsilon^t_q,\epsilon^r_g)=(0.5,0.13)$, which have been obtained in Ref.~\cite{FerreiradeLima:2016gcz} under similar kinematical conditions. Throughout, we include the expected background distributions and the relative reduction of the SM $\mathrm{jet}+\gamma$ production after quark/gluon tagging. For the considered mass range of Eq.~\eqref{eq:massselection}, the SM continuum production is dominated by processes with a final state quark and the reduction of background cross section is mostly determined by the tagger's working point. Compared to the jet$+\gamma$ production cross section after selection cuts of $\sim 6.6~\text{pb}$, quark/gluon tagging reduces the cross section by $\sim 47\%$. In Fig.~\ref{fig:tagb}, we add the tagged distributions as a separate category to the likelihood, which can be compared to the ``raw'' $\cos\theta^\ast$ discrimination in Fig.~\ref{fig:taga}. As can be seen, some of the competing distributions can be excluded to support statistical preference for one particular model for moderate luminosities. This simple hypothesis test which indicates statistical preference between two well-defined hypotheses does not constitute a coupling measurement, but will be the first step in this direction (experimental results related to the Higgs~can be found in e.g.~\cite{Aad:2013xqa}).
 
\begin{wrapfigure}[15]{r}{6.7cm}	
\hfill\parbox{6.5cm}{
\centering
 \includegraphics[width=6.5cm]{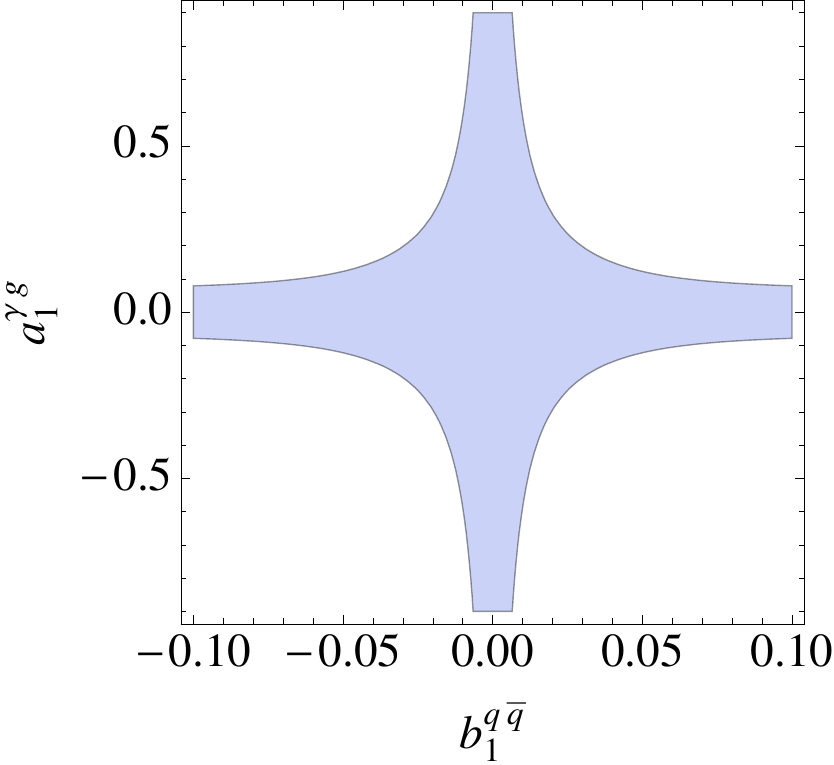}
 \caption{\label{fig:jetaextr} Coupling region from a toy Monte Carlo analysis at $\sqrt{s}=13~\text{TeV}$ for signal cross sections $\sigma<23.4~\text{fb}$, the estimated $5\sigma$ discovery at ${\cal{L}}=300~\text{fb}^{-1}$. For details see text.}}
\end{wrapfigure}

Continuing with this particular example, at higher luminosity, once the $b^{q\bar q}_1 \otimes a^{\gamma g}_1$ character is established, the overall rate in excess the background measurement can be used to constrain the couplings of the model. We show this representatively in Fig.~\ref{fig:jetaextr} for our $M=500~\text{GeV}$ benchmark, showing all parameter combinations that yield a maximum signal cross section of the estimated 5$\sigma$ discovery cross section of $\sim 23.4~\text{fb}$. This MC-based toy extraction also demonstrates that additional information from di-jet measurements is necessary to avoid blind directions, which arise from fitting the narrow width approximation: For large values of e.g. $b^{q\bar q}_1$, the jet$+\gamma$ signal cross section scales as function of $a^{\gamma g}_1$ alone and the limit is saturated by a constraint on this single coupling. 
A di-jet constraint, on the other hand, will close the blind direction for large values of $b^{q\bar q}_1$. A remaining question is whether the estimated 5$\sigma$ discovery cross section of $\sim 23.4~\text{fb}$ corresponds to a choice of parameters for which our main assumption, i.e. the narrow width approximation is  still valid.
Scanning the couplings inputing this discovery threshold as a constraint, we can obtain $b^{q\bar q}_1$ as a function of $a^{\gamma g}_1$, which then allows to express $\Gamma_X$ as function of $b^{q\bar q}_1$ alone. For our particular benchmark we obtain $\Gamma_X/M\lesssim 0.03$ for $|b^{q\bar q}_1|<1$, highlighting the good validity of the narrow width approximation in this case. The branching ratios for the parameter choices that are allowed this way vary between dominant $j\gamma$-like final states for $|b^{q\bar q}_1|\ll 1$ to di--jet like final states for $|b^{q\bar q}_1|\sim1$, which again shows the necessity of studying complementary channels in case of a discovery in final states as described in this work.

\section{Summary} 
The absence of conclusive hints for new interactions beyond the Standard Model motivates a wider approach to searches for new states that fall within the energy coverage of the LHC or future hadron colliders. Scenarios of QCD-charged new states that could arise in a range of composite Higgs models have been less investigated in the past, 
and this work presents a detailed investigation of possible $2\to 2$ scattering processes of such states with jet$+\gamma$ and jet$+Z$ production. This provides the theoretical underpinning for future searches, generalizing the current signal modelling of ATLAS and CMS~\cite{Aad:2015ywd,CMS:2016qtb,Aaboud:2017nak,CMS:2017eej}. We have based our analysis around a decomposition of the scattering amplitudes into irreducible categories, thereby widening the phenomenological range beyond the constraints of effective field theory. This comes at the price of a strict on-shell formulation of the hard scattering amplitude, which removes a straightforward application of CP-discriminating techniques. However, the phenomenology of the scattering amplitudes proves rich and is observable at the LHC over a broad particle mass range. Naturally, the lack of any observation so far that could be interpreted along the lines of this paper leaves the parameter space vastly bigger than the humble set of angular and kinematic distributions of $2\to 2$ scattering is capable to constrain. However, we have shown that by adapting angles from the Higgs characterization program, at least partial insights can be gained into the spin of such a produced state. In particular, we have also demonstrated how advances in jet technology (specifically quark/gluon tagging) can assist the resonance's discovery or spectroscopy in the future.

\subsection*{Acknowledgements}
C.E. is supported by the IPPP Associateship scheme and by the UK Science and Technology Facilities Council (STFC) under grant ST/P000746/1.
G.F. would like to thank C.~Petersson for discussions.
M.S. is supported in part by the European Commission through the ``HiggsTools'' Initial Training Network PITN-GA-2012-316704. 


\providecommand{\href}[2]{#2}\begingroup\raggedright\endgroup


\begin{thebibliography}{10}

\bibitem{Kilic:2009mi}
C.~Kilic, T.~Okui and R.~Sundrum, \emph{{Vectorlike Confinement at the LHC}},
  \href{http://dx.doi.org/10.1007/JHEP02(2010)018}{\emph{JHEP} {\bfseries 02}
  (2010) 018}, [\href{https://arxiv.org/abs/0906.0577}{{\ttfamily 0906.0577}}].

\bibitem{Schumann:2011ji}
S.~Schumann, A.~Renaud and D.~Zerwas, \emph{{Hadronically decaying
  color-adjoint scalars at the LHC}},
  \href{http://dx.doi.org/10.1007/JHEP09(2011)074}{\emph{JHEP} {\bfseries 09}
  (2011) 074}, [\href{https://arxiv.org/abs/1108.2957}{{\ttfamily 1108.2957}}].

\bibitem{Hayot:1980gg}
F.~Hayot and O.~Napoly, \emph{{Detecting a Heavy Colored Object at the Fnal
  Tevatron}}, \href{http://dx.doi.org/10.1007/BF01436311}{\emph{Z. Phys.}
  {\bfseries C7} (1981) 229}.

\bibitem{Ellis:1980hz}
J.~R. Ellis, M.~K. Gaillard, D.~V. Nanopoulos and P.~Sikivie, \emph{{Can One
  Tell Technicolor from a Hole in the Ground?}},
  \href{http://dx.doi.org/10.1016/0550-3213(81)90133-4}{\emph{Nucl. Phys.}
  {\bfseries B182} (1981) 529--545}.

\bibitem{Belyaev:1999xe}
A.~Belyaev, R.~Rosenfeld and A.~R. Zerwekh, \emph{{Tevatron potential for
  technicolor search with prompt photons}},
  \href{http://dx.doi.org/10.1016/S0370-2693(99)00869-2}{\emph{Phys. Lett.}
  {\bfseries B462} (1999) 150--157},
  [\href{https://arxiv.org/abs/hep-ph/9905468}{{\ttfamily hep-ph/9905468}}].

\bibitem{Bai:2010mn}
Y.~Bai and A.~Martin, \emph{{Topological Pions}},
  \href{http://dx.doi.org/10.1016/j.physletb.2010.08.058}{\emph{Phys. Lett.}
  {\bfseries B693} (2010) 292--295},
  [\href{https://arxiv.org/abs/1003.3006}{{\ttfamily 1003.3006}}].

\bibitem{Cacciapaglia:2015eqa}
G.~Cacciapaglia, H.~Cai, A.~Deandrea, T.~Flacke, S.~J. Lee and A.~Parolini,
  \emph{{Composite scalars at the LHC: the Higgs, the Sextet and the Octet}},
  \href{http://dx.doi.org/10.1007/JHEP11(2015)201}{\emph{JHEP} {\bfseries 11}
  (2015) 201}, [\href{https://arxiv.org/abs/1507.02283}{{\ttfamily
  1507.02283}}].

\bibitem{Bai:2016czm}
Y.~Bai, V.~Barger and J.~Berger, \emph{{Constraints on color-octet companions
  of a 750 GeV heavy pion from dijet and photon plus jet resonance searches}},
  \href{http://dx.doi.org/10.1103/PhysRevD.94.011701}{\emph{Phys. Rev.}
  {\bfseries D94} (2016) 011701},
  [\href{https://arxiv.org/abs/1604.07835}{{\ttfamily 1604.07835}}].

\bibitem{Bizot:2016zyu}
N.~Bizot, M.~Frigerio, M.~Knecht and J.-L. Kneur, \emph{{Nonperturbative
  analysis of the spectrum of meson resonances in an ultraviolet-complete
  composite-Higgs model}},
  \href{http://dx.doi.org/10.1103/PhysRevD.95.075006}{\emph{Phys. Rev.}
  {\bfseries D95} (2017) 075006},
  [\href{https://arxiv.org/abs/1610.09293}{{\ttfamily 1610.09293}}].

\bibitem{Hayreter:2017wra}
A.~Hayreter and G.~Valencia, \emph{{LHC constraints on color octet scalars}},
  \href{https://arxiv.org/abs/1703.04164}{{\ttfamily 1703.04164}}.

\bibitem{Barnard:2013zea}
J.~Barnard, T.~Gherghetta and T.~S. Ray, \emph{{UV descriptions of composite
  Higgs models without elementary scalars}},
  \href{http://dx.doi.org/10.1007/JHEP02(2014)002}{\emph{JHEP} {\bfseries 02}
  (2014) 002}, [\href{https://arxiv.org/abs/1311.6562}{{\ttfamily 1311.6562}}].

\bibitem{Ferretti:2013kya}
G.~Ferretti and D.~Karateev, \emph{{Fermionic UV completions of Composite Higgs
  models}}, \href{http://dx.doi.org/10.1007/JHEP03(2014)077}{\emph{JHEP}
  {\bfseries 03} (2014) 077},
  [\href{https://arxiv.org/abs/1312.5330}{{\ttfamily 1312.5330}}].

\bibitem{Ferretti:2014qta}
G.~Ferretti, \emph{{UV Completions of Partial Compositeness: The Case for a
  SU(4) Gauge Group}},
  \href{http://dx.doi.org/10.1007/JHEP06(2014)142}{\emph{JHEP} {\bfseries 06}
  (2014) 142}, [\href{https://arxiv.org/abs/1404.7137}{{\ttfamily 1404.7137}}].

\bibitem{Vecchi:2015fma}
L.~Vecchi, \emph{{A dangerous irrelevant UV-completion of the composite
  Higgs}}, \href{http://dx.doi.org/10.1007/JHEP02(2017)094}{\emph{JHEP}
  {\bfseries 02} (2017) 094},
  [\href{https://arxiv.org/abs/1506.00623}{{\ttfamily 1506.00623}}].

\bibitem{Ferretti:2016upr}
G.~Ferretti, \emph{{Gauge theories of Partial Compositeness: Scenarios for
  Run-II of the LHC}},
  \href{http://dx.doi.org/10.1007/JHEP06(2016)107}{\emph{JHEP} {\bfseries 06}
  (2016) 107}, [\href{https://arxiv.org/abs/1604.06467}{{\ttfamily
  1604.06467}}].

\bibitem{Belyaev:2016ftv}
A.~Belyaev, G.~Cacciapaglia, H.~Cai, G.~Ferretti, T.~Flacke, A.~Parolini
  et~al., \emph{{Di-boson signatures as Standard Candles for Partial
  Compositeness}}, \href{http://dx.doi.org/10.1007/JHEP01(2017)094}{\emph{JHEP}
  {\bfseries 01} (2017) 094},
  [\href{https://arxiv.org/abs/1610.06591}{{\ttfamily 1610.06591}}].

\bibitem{Cabibbo:1983bk}
N.~Cabibbo, L.~Maiani and Y.~Srivastava, \emph{{Anomalous Z Decays: Excited
  Leptons?}}, \href{http://dx.doi.org/10.1016/0370-2693(84)91850-1}{\emph{Phys.
  Lett.} {\bfseries B139} (1984) 459--463}.

\bibitem{DeRujula:1983ak}
A.~De~Rujula, L.~Maiani and R.~Petronzio, \emph{{Search for Excited Quarks}},
  \href{http://dx.doi.org/10.1016/0370-2693(84)90930-4}{\emph{Phys. Lett.}
  {\bfseries 140B} (1984) 253--258}.

\bibitem{Kleiss:1987ab}
R.~Kleiss and P.~M. Zerwas, \emph{{EXCITED QUARKS AND LEPTONS AT e+ e- AND p p
  COLLIDERS}},  in \emph{{Workshop on Physics at Future Accelerators La Thuile,
  Italy and Geneva, Switz., January 7-13, 1987}}, 1987.

\bibitem{Baur:1987ga}
U.~Baur, I.~Hinchliffe and D.~Zeppenfeld, \emph{{Excited Quark Production at
  Hadron Colliders}},
  \href{http://dx.doi.org/10.1142/S0217751X87000661}{\emph{Int. J. Mod. Phys.}
  {\bfseries A2} (1987) 1285}.

\bibitem{Baur:1989kv}
U.~Baur, M.~Spira and P.~M. Zerwas, \emph{{Excited Quark and Lepton Production
  at Hadron Colliders}},
  \href{http://dx.doi.org/10.1103/PhysRevD.42.815}{\emph{Phys. Rev.} {\bfseries
  D42} (1990) 815--824}.

\bibitem{Bhattacharya:2009xg}
S.~Bhattacharya, S.~S. Chauhan, B.~C. Choudhary and D.~Choudhury, \emph{{Quark
  Excitations Through the Prism of Direct Photon Plus Jet at the LHC}},
  \href{http://dx.doi.org/10.1103/PhysRevD.80.015014}{\emph{Phys. Rev.}
  {\bfseries D80} (2009) 015014},
  [\href{https://arxiv.org/abs/0901.3927}{{\ttfamily 0901.3927}}].

\bibitem{Choi:2002jk}
S.~Y. Choi, D.~J. Miller, M.~M. Muhlleitner and P.~M. Zerwas,
  \emph{{Identifying the Higgs spin and parity in decays to Z pairs}},
  \href{http://dx.doi.org/10.1016/S0370-2693(02)03191-X}{\emph{Phys. Lett.}
  {\bfseries B553} (2003) 61--71},
  [\href{https://arxiv.org/abs/hep-ph/0210077}{{\ttfamily hep-ph/0210077}}].

\bibitem{Gao:2010qx}
Y.~Gao, A.~V. Gritsan, Z.~Guo, K.~Melnikov, M.~Schulze and N.~V. Tran,
  \emph{{Spin determination of single-produced resonances at hadron
  colliders}}, \href{http://dx.doi.org/10.1103/PhysRevD.81.075022}{\emph{Phys.
  Rev.} {\bfseries D81} (2010) 075022},
  [\href{https://arxiv.org/abs/1001.3396}{{\ttfamily 1001.3396}}].

\bibitem{DeRujula:2010ys}
A.~De~Rujula, J.~Lykken, M.~Pierini, C.~Rogan and M.~Spiropulu, \emph{{Higgs
  look-alikes at the LHC}},
  \href{http://dx.doi.org/10.1103/PhysRevD.82.013003}{\emph{Phys. Rev.}
  {\bfseries D82} (2010) 013003},
  [\href{https://arxiv.org/abs/1001.5300}{{\ttfamily 1001.5300}}].

\bibitem{Bolognesi:2012mm}
S.~Bolognesi, Y.~Gao, A.~V. Gritsan, K.~Melnikov, M.~Schulze, N.~V. Tran
  et~al., \emph{{On the spin and parity of a single-produced resonance at the
  LHC}}, \href{http://dx.doi.org/10.1103/PhysRevD.86.095031}{\emph{Phys. Rev.}
  {\bfseries D86} (2012) 095031},
  [\href{https://arxiv.org/abs/1208.4018}{{\ttfamily 1208.4018}}].

\bibitem{Khachatryan:2014kca}
{\scshape CMS} collaboration, V.~Khachatryan et~al., \emph{{Constraints on the
  spin-parity and anomalous HVV couplings of the Higgs boson in proton
  collisions at 7 and 8 TeV}},
  \href{http://dx.doi.org/10.1103/PhysRevD.92.012004}{\emph{Phys. Rev.}
  {\bfseries D92} (2015) 012004},
  [\href{https://arxiv.org/abs/1411.3441}{{\ttfamily 1411.3441}}].

\bibitem{Miller:2001bi}
D.~J. Miller, S.~Y. Choi, B.~Eberle, M.~M. Muhlleitner and P.~M. Zerwas,
  \emph{{Measuring the spin of the Higgs boson}},
  \href{http://dx.doi.org/10.1016/S0370-2693(01)00317-3}{\emph{Phys. Lett.}
  {\bfseries B505} (2001) 149--154},
  [\href{https://arxiv.org/abs/hep-ph/0102023}{{\ttfamily hep-ph/0102023}}].

\bibitem{Buszello:2002uu}
C.~P. Buszello, I.~Fleck, P.~Marquard and J.~J. van~der Bij, \emph{{Prospective
  analysis of spin- and CP-sensitive variables in H $\to$ Z Z $\to$ l(1)$^+$
  l(1)$^-$ l(2)$^+$ l(2)$^-$ at the LHC}},
  \href{http://dx.doi.org/10.1140/epjc/s2003-01392-0}{\emph{Eur. Phys. J.}
  {\bfseries C32} (2004) 209--219},
  [\href{https://arxiv.org/abs/hep-ph/0212396}{{\ttfamily hep-ph/0212396}}].

\bibitem{Choi:2012yg}
S.~Y. Choi, M.~M. Muhlleitner and P.~M. Zerwas, \emph{{Theoretical Basis of
  Higgs-Spin Analysis in $H \to \gamma\gamma$ and $Z\gamma$ Decays}},
  \href{http://dx.doi.org/10.1016/j.physletb.2012.11.050}{\emph{Phys. Lett.}
  {\bfseries B718} (2013) 1031--1035},
  [\href{https://arxiv.org/abs/1209.5268}{{\ttfamily 1209.5268}}].

\bibitem{Jacob:1959at}
M.~Jacob and G.~C. Wick, \emph{{On the general theory of collisions for
  particles with spin}},
  \href{http://dx.doi.org/10.1016/0003-4916(59)90051-X}{\emph{Annals Phys.}
  {\bfseries 7} (1959) 404--428}.

\bibitem{Aad:2015ywd}
{\scshape ATLAS} collaboration, G.~Aad et~al., \emph{{Search for new phenomena
  with photon+jet events in proton-proton collisions at $ \sqrt{s}=13 $ TeV
  with the ATLAS detector}},
  \href{http://dx.doi.org/10.1007/JHEP03(2016)041}{\emph{JHEP} {\bfseries 03}
  (2016) 041}, [\href{https://arxiv.org/abs/1512.05910}{{\ttfamily
  1512.05910}}].

\bibitem{CMS:2016qtb}
{\scshape CMS} collaboration, \emph{{Search for excited
  quarks in the photon + jet final state in proton proton collisions at 13
  TeV}}, CMS-PAS-EXO-16-015.

\bibitem{Aaboud:2017nak}
{\scshape ATLAS} collaboration, M.~Aaboud et~al., \emph{{Search for new
  phenomena in high-mass final states with a photon and a jet from $pp$
  collisions at $\sqrt{s}$ = 13 TeV with the ATLAS detector}},
  \href{https://arxiv.org/abs/1709.10440}{{\ttfamily 1709.10440}}.

\bibitem{CMS:2017eej}
{\scshape CMS} collaboration, \emph{{Search for excited
  states of light and heavy flavor quarks in the $\gamma\mathrm{+jet}$ final
  state in proton-proton collisions at $\sqrt{s}=13~\mathrm{TeV}$}}, CMS-PAS-EXO-17-002.

\bibitem{Cacciari:2015ela}
M.~Cacciari, L.~Del~Debbio, J.~R. Espinosa, A.~D. Polosa and M.~Testa, \emph{{A
  note on the fate of the Landau–Yang theorem in non-Abelian gauge
  theories}},
  \href{http://dx.doi.org/10.1016/j.physletb.2015.12.053}{\emph{Phys. Lett.}
  {\bfseries B753} (2016) 476--481},
  [\href{https://arxiv.org/abs/1509.07853}{{\ttfamily 1509.07853}}].

\bibitem{Panico:2016ary}
G.~Panico, L.~Vecchi and A.~Wulzer, \emph{{Resonant Diphoton Phenomenology
  Simplified}}, \href{http://dx.doi.org/10.1007/JHEP06(2016)184}{\emph{JHEP}
  {\bfseries 06} (2016) 184},
  [\href{https://arxiv.org/abs/1603.04248}{{\ttfamily 1603.04248}}].

\bibitem{Alwall:2011uj}
J.~Alwall, M.~Herquet, F.~Maltoni, O.~Mattelaer and T.~Stelzer, \emph{{MadGraph
  5 : Going Beyond}},
  \href{http://dx.doi.org/10.1007/JHEP06(2011)128}{\emph{JHEP} {\bfseries 06}
  (2011) 128}, [\href{https://arxiv.org/abs/1106.0522}{{\ttfamily 1106.0522}}].

\bibitem{Alwall:2014hca}
J.~Alwall, R.~Frederix, S.~Frixione, V.~Hirschi, F.~Maltoni, O.~Mattelaer
  et~al., \emph{{The automated computation of tree-level and next-to-leading
  order differential cross sections, and their matching to parton shower
  simulations}}, \href{http://dx.doi.org/10.1007/JHEP07(2014)079}{\emph{JHEP}
  {\bfseries 07} (2014) 079},
  [\href{https://arxiv.org/abs/1405.0301}{{\ttfamily 1405.0301}}].

\bibitem{Murayama:1992gi}
H.~Murayama, I.~Watanabe and K.~Hagiwara, \emph{{HELAS: HELicity amplitude
  subroutines for Feynman diagram evaluations}}, KEK-91-11 (1992).

\bibitem{Alloul:2013bka}
A.~Alloul, N.~D. Christensen, C.~Degrande, C.~Duhr and B.~Fuks,
  \emph{{FeynRules 2.0 - A complete toolbox for tree-level phenomenology}},
  \href{http://dx.doi.org/10.1016/j.cpc.2014.04.012}{\emph{Comput. Phys.
  Commun.} {\bfseries 185} (2014) 2250--2300},
  [\href{https://arxiv.org/abs/1310.1921}{{\ttfamily 1310.1921}}].

\bibitem{Degrande:2011ua}
C.~Degrande, C.~Duhr, B.~Fuks, D.~Grellscheid, O.~Mattelaer and T.~Reiter,
  \emph{{UFO - The Universal FeynRules Output}},
  \href{http://dx.doi.org/10.1016/j.cpc.2012.01.022}{\emph{Comput. Phys.
  Commun.} {\bfseries 183} (2012) 1201--1214},
  [\href{https://arxiv.org/abs/1108.2040}{{\ttfamily 1108.2040}}].

\bibitem{Bahr:2008pv}
M.~Bahr et~al., \emph{{Herwig++ Physics and Manual}},
  \href{http://dx.doi.org/10.1140/epjc/s10052-008-0798-9}{\emph{Eur. Phys. J.}
  {\bfseries C58} (2008) 639--707},
  [\href{https://arxiv.org/abs/0803.0883}{{\ttfamily 0803.0883}}].

\bibitem{Bellm:2015jjp}
J.~Bellm et~al., \emph{{Herwig 7.0/Herwig++ 3.0 release note}},
  \href{http://dx.doi.org/10.1140/epjc/s10052-016-4018-8}{\emph{Eur. Phys. J.}
  {\bfseries C76} (2016) 196},
  [\href{https://arxiv.org/abs/1512.01178}{{\ttfamily 1512.01178}}].

\bibitem{Plehn:2001nj}
T.~Plehn, D.~L. Rainwater and D.~Zeppenfeld, \emph{{Determining the structure
  of Higgs couplings at the LHC}},
  \href{http://dx.doi.org/10.1103/PhysRevLett.88.051801}{\emph{Phys. Rev.
  Lett.} {\bfseries 88} (2002) 051801},
  [\href{https://arxiv.org/abs/hep-ph/0105325}{{\ttfamily hep-ph/0105325}}].

\bibitem{Hankele:2006ma}
V.~Hankele, G.~Klamke, D.~Zeppenfeld and T.~Figy, \emph{{Anomalous Higgs boson
  couplings in vector boson fusion at the CERN LHC}},
  \href{http://dx.doi.org/10.1103/PhysRevD.74.095001}{\emph{Phys. Rev.}
  {\bfseries D74} (2006) 095001},
  [\href{https://arxiv.org/abs/hep-ph/0609075}{{\ttfamily hep-ph/0609075}}].

\bibitem{Klamke:2007cu}
G.~Klamke and D.~Zeppenfeld, \emph{{Higgs plus two jet production via gluon
  fusion as a signal at the CERN LHC}},
  \href{http://dx.doi.org/10.1088/1126-6708/2007/04/052}{\emph{JHEP} {\bfseries
  04} (2007) 052}, [\href{https://arxiv.org/abs/hep-ph/0703202}{{\ttfamily
  hep-ph/0703202}}].

\bibitem{Hagiwara:2009wt}
K.~Hagiwara, Q.~Li and K.~Mawatari, \emph{{Jet angular correlation in
  vector-boson fusion processes at hadron colliders}},
  \href{http://dx.doi.org/10.1088/1126-6708/2009/07/101}{\emph{JHEP} {\bfseries
  07} (2009) 101}, [\href{https://arxiv.org/abs/0905.4314}{{\ttfamily
  0905.4314}}].

\bibitem{Englert:2012ct}
C.~Englert, M.~Spannowsky and M.~Takeuchi, \emph{{Measuring Higgs CP and
  couplings with hadronic event shapes}},
  \href{http://dx.doi.org/10.1007/JHEP06(2012)108}{\emph{JHEP} {\bfseries 06}
  (2012) 108}, [\href{https://arxiv.org/abs/1203.5788}{{\ttfamily 1203.5788}}].

\bibitem{Englert:2012xt}
C.~Englert, D.~Goncalves-Netto, K.~Mawatari and T.~Plehn, \emph{{Higgs Quantum
  Numbers in Weak Boson Fusion}},
  \href{http://dx.doi.org/10.1007/JHEP01(2013)148}{\emph{JHEP} {\bfseries 01}
  (2013) 148}, [\href{https://arxiv.org/abs/1212.0843}{{\ttfamily 1212.0843}}].

\bibitem{Dolan:2014upa}
M.~J. Dolan, P.~Harris, M.~Jankowiak and M.~Spannowsky, \emph{{Constraining
  $CP$-violating Higgs Sectors at the LHC using gluon fusion}},
  \href{http://dx.doi.org/10.1103/PhysRevD.90.073008}{\emph{Phys. Rev.}
  {\bfseries D90} (2014) 073008},
  [\href{https://arxiv.org/abs/1406.3322}{{\ttfamily 1406.3322}}].

\bibitem{Sirunyan:2016iap}
{\scshape CMS} collaboration, A.~M. Sirunyan et~al., \emph{{Search for dijet
  resonances in proton–proton collisions at $\sqrt{s}$ = 13 TeV and
  constraints on dark matter and other models}},
  \href{http://dx.doi.org/10.1016/j.physletb.2017.09.029,
  10.1016/j.physletb.2017.02.012}{\emph{Phys. Lett.} {\bfseries B769} (2017)
  520--542}, [\href{https://arxiv.org/abs/1611.03568}{{\ttfamily 1611.03568}}].

\bibitem{Aaboud:2017yvp}
{\scshape ATLAS} collaboration, M.~Aaboud et~al., \emph{{Search for new
  phenomena in dijet events using 37 fb$^{-1}$ of $pp$ collision data collected
  at $\sqrt{s}=$13 TeV with the ATLAS detector}},
  \href{http://dx.doi.org/10.1103/PhysRevD.96.052004}{\emph{Phys. Rev.}
  {\bfseries D96} (2017) 052004},
  [\href{https://arxiv.org/abs/1703.09127}{{\ttfamily 1703.09127}}].

\bibitem{ATLAS:2012ds}
{\scshape ATLAS} collaboration, G.~Aad et~al., \emph{{Search for pair-produced
  massive coloured scalars in four-jet final states with the ATLAS detector in
  proton-proton collisions at $\sqrt{s}=7$ TeV}},
  \href{http://dx.doi.org/10.1140/epjc/s10052-012-2263-z}{\emph{Eur. Phys. J.}
  {\bfseries C73} (2013) 2263},
  [\href{https://arxiv.org/abs/1210.4826}{{\ttfamily 1210.4826}}].

\bibitem{Khachatryan:2014lpa}
{\scshape CMS} collaboration, V.~Khachatryan et~al., \emph{{Search for
  pair-produced resonances decaying to jet pairs in proton–proton collisions
  at $\sqrt{s} =$ 8 TeV}},
  \href{http://dx.doi.org/10.1016/j.physletb.2015.04.045}{\emph{Phys. Lett.}
  {\bfseries B747} (2015) 98--119},
  [\href{https://arxiv.org/abs/1412.7706}{{\ttfamily 1412.7706}}].

\bibitem{Cabibbo:1965zzb}
N.~Cabibbo and A.~Maksymowicz, \emph{{Angular Correlations in Ke-4 Decays and
  Determination of Low-Energy pi-pi Phase Shifts}},
  \href{http://dx.doi.org/10.1103/PhysRev.137.B438,
  10.1103/PhysRev.168.1926}{\emph{Phys. Rev.} {\bfseries 137} (1965)
  B438--B443}.

\bibitem{Trueman:1978kh}
T.~L. Trueman, \emph{{$\phi \phi$ Decay as a Parity and Signature Test}},
  \href{http://dx.doi.org/10.1103/PhysRevD.18.3423}{\emph{Phys. Rev.}
  {\bfseries D18} (1978) 3423}.

\bibitem{Collins:1977iv}
J.~C. Collins and D.~E. Soper, \emph{{Angular Distribution of Dileptons in
  High-Energy Hadron Collisions}},
  \href{http://dx.doi.org/10.1103/PhysRevD.16.2219}{\emph{Phys. Rev.}
  {\bfseries D16} (1977) 2219}.

\bibitem{DellAquila:1985mtb}
J.~R. Dell'Aquila and C.~A. Nelson, \emph{{$P$ or {CP} Determination by
  Sequential Decays: V1 V2 Modes With Decays Into $\bar{\ell}$epton (A) $\ell$
  (B) And/or $\bar{q}$ (A) $q $ (B)}},
  \href{http://dx.doi.org/10.1103/PhysRevD.33.80}{\emph{Phys. Rev.} {\bfseries
  D33} (1986) 80}.

\bibitem{DellAquila:1985jin}
J.~R. Dell'Aquila and C.~A. Nelson, \emph{{Distinguishing a Spin 0 Technipion
  and an Elementary Higgs Boson: V1 V2 Modes With Decays Into $\bar{\ell}$epton
  (A) $\ell (B$) And/or $\bar{q}$ (A) $q (B$)}},
  \href{http://dx.doi.org/10.1103/PhysRevD.33.93}{\emph{Phys. Rev.} {\bfseries
  D33} (1986) 93}.

\bibitem{Nelson:1986ki}
C.~A. Nelson, \emph{{Correlation Between Decay Planes in Higgs Boson Decays
  Into $W$ Pair (Into $Z$ Pair)}},
  \href{http://dx.doi.org/10.1103/PhysRevD.37.1220}{\emph{Phys. Rev.}
  {\bfseries D37} (1988) 1220}.

\bibitem{Englert:2010ud}
C.~Englert, C.~Hackstein and M.~Spannowsky, \emph{{Measuring spin and CP from
  semi-hadronic ZZ decays using jet substructure}},
  \href{http://dx.doi.org/10.1103/PhysRevD.82.114024}{\emph{Phys. Rev.}
  {\bfseries D82} (2010) 114024},
  [\href{https://arxiv.org/abs/1010.0676}{{\ttfamily 1010.0676}}].

\bibitem{ATLAS:2016wzt}
{\scshape ATLAS} collaboration, \emph{{Discrimination of
  Light Quark and Gluon Jets in $pp$ collisions at $\sqrt{s} = 8$ TeV with the
  ATLAS Detector}}, ATLAS-CONF-2016-034.

\bibitem{CMS:2017wyc}
{\scshape CMS} collaboration, \emph{{Jet algorithms
  performance in 13 TeV data}}, CMS-PAS-JME-16-003.

\bibitem{Gallicchio:2012ez}
J.~Gallicchio and M.~D. Schwartz, \emph{{Quark and Gluon Jet Substructure}},
  \href{http://dx.doi.org/10.1007/JHEP04(2013)090}{\emph{JHEP} {\bfseries 04}
  (2013) 090}, [\href{https://arxiv.org/abs/1211.7038}{{\ttfamily 1211.7038}}].

\bibitem{FerreiradeLima:2016gcz}
D.~Ferreira~de Lima, P.~Petrov, D.~Soper and M.~Spannowsky, \emph{{Quark-Gluon
  tagging with Shower Deconstruction: Unearthing dark matter and Higgs
  couplings}}, \href{http://dx.doi.org/10.1103/PhysRevD.95.034001}{\emph{Phys.
  Rev.} {\bfseries D95} (2017) 034001},
  [\href{https://arxiv.org/abs/1607.06031}{{\ttfamily 1607.06031}}].

\bibitem{ATL-COM-DAQ-2017-117}
{\scshape ATLAS} collaboration, \emph{{Trigger Menu in 2016}}, ATL-DAQ-PUB-2017-001.

\bibitem{Cacciari:2008gp}
M.~Cacciari, G.~P. Salam and G.~Soyez, \emph{{The Anti-k(t) jet clustering
  algorithm}},
  \href{http://dx.doi.org/10.1088/1126-6708/2008/04/063}{\emph{JHEP} {\bfseries
  04} (2008) 063}, [\href{https://arxiv.org/abs/0802.1189}{{\ttfamily
  0802.1189}}].

\bibitem{Cacciari:2011ma}
M.~Cacciari, G.~P. Salam and G.~Soyez, \emph{{FastJet User Manual}},
  \href{http://dx.doi.org/10.1140/epjc/s10052-012-1896-2}{\emph{Eur. Phys. J.}
  {\bfseries C72} (2012) 1896},
  [\href{https://arxiv.org/abs/1111.6097}{{\ttfamily 1111.6097}}].

\bibitem{Tapper:2013yva}
{\scshape CMS} collaboration, A.~Tapper and D.~Acosta, \emph{{CMS Technical
  Design Report for the Level-1 Trigger Upgrade}}, CERN-LHCC-2013-011, CMS-TDR-12, CMS-TDR-012.

\bibitem{deFavereau:2013fsa}
{\scshape DELPHES 3} collaboration, J.~de~Favereau, C.~Delaere, P.~Demin,
  A.~Giammanco, V.~Lemaitre, A.~Mertens et~al., \emph{{DELPHES 3, A modular
  framework for fast simulation of a generic collider experiment}},
  \href{http://dx.doi.org/10.1007/JHEP02(2014)057}{\emph{JHEP} {\bfseries 02}
  (2014) 057}, [\href{https://arxiv.org/abs/1307.6346}{{\ttfamily 1307.6346}}].

\bibitem{Gras:2017jty}
P.~Gras, S.~Hoeche, D.~Kar, A.~Larkoski, L.~L\''onnblad, S.~Platzer et~al.,
  \emph{{Systematics of quark/gluon tagging}},
  \href{https://arxiv.org/abs/1704.03878}{{\ttfamily 1704.03878}}.

\bibitem{Junk:1999kv}
T.~Junk, \emph{{Confidence level computation for combining searches with small
  statistics}},
  \href{http://dx.doi.org/10.1016/S0168-9002(99)00498-2}{\emph{Nucl. Instrum.
  Meth.} {\bfseries A434} (1999) 435--443},
  [\href{https://arxiv.org/abs/hep-ex/9902006}{{\ttfamily hep-ex/9902006}}].

\bibitem{Read:2000ru}
A.~L. Read, \emph{{Modified frequentist analysis of search results (The CL(s)
  method)}},  in \emph{{Workshop on confidence limits, CERN, Geneva,
  Switzerland, 17-18 Jan 2000: Proceedings}}, pp.~81--101, 2000.

\bibitem{Read:2002hq}
A.~L. Read, \emph{{Presentation of search results: The CL(s) technique}},
  \href{http://dx.doi.org/10.1088/0954-3899/28/10/313}{\emph{J. Phys.}
  {\bfseries G28} (2002) 2693--2704}.

\bibitem{Aad:2013xqa}
{\scshape ATLAS} collaboration, G.~Aad et~al., \emph{{Evidence for the spin-0
  nature of the Higgs boson using ATLAS data}},
  \href{http://dx.doi.org/10.1016/j.physletb.2013.08.026}{\emph{Phys. Lett.}
  {\bfseries B726} (2013) 120--144},
  [\href{https://arxiv.org/abs/1307.1432}{{\ttfamily 1307.1432}}].

\end{thebibliography}

\end{document}